\documentclass[iop]{emulateapj}
\usepackage{multirow}
\usepackage[english]{babel}
\usepackage{natbib}
\usepackage[usenames,dvipsnames]{color}
\usepackage{rotating}
\usepackage{graphicx}
\usepackage{ulem} 

\usepackage{url} 

\newcommand{\myarcsec}{\hbox{$.\!\!^{\prime\prime}$}}
\newcommand{\myarcmin}{\hbox{$.\!\!^{\prime}$}}

\usepackage{hyperref}
\usepackage{breakurl} 

\slugcomment{ApJS, accepted}
\shorttitle{NGVS photometric redshifts}
\shortauthors{Raichoor et al.}

\begin{document}

\title{The Next Generation Virgo Cluster Survey. XV. The photometric redshift estimation for background sources}

\author{
A.~Raichoor\altaffilmark{1},
S.~Mei\altaffilmark{1,2},
T.~Erben\altaffilmark{3},
H.~Hildebrandt\altaffilmark{3},
M.~Huertas-Company\altaffilmark{1,2},
O.~Ilbert\altaffilmark{4},
R.~Licitra\altaffilmark{1,2},
N.M.~Ball\altaffilmark{5,6},
S.~Boissier\altaffilmark{4,7},
A.~Boselli\altaffilmark{4},
Y.-T.~Chen\altaffilmark{8},
P.~C\^{o}t\'{e}\altaffilmark{5},
J.-C.~Cuillandre\altaffilmark{9},
P.A.~Duc\altaffilmark{10},
P.R.~Durrell\altaffilmark{11},
L.~Ferrarese\altaffilmark{5},
P.~Guhathakurta\altaffilmark{12},
S.D.J.~Gwyn\altaffilmark{5},
J.J.~Kavelaars\altaffilmark{5},
A.~Lan\c{c}on\altaffilmark{13},
C.~Liu\altaffilmark{14,15},
L.A.~MacArthur\altaffilmark{5},
M.~Muller\altaffilmark{12},
R.P.~Mu\~noz\altaffilmark{16,13},
E.W.~Peng\altaffilmark{17,18},
T.H.~Puzia\altaffilmark{16},
M.~Sawicki\altaffilmark{19},
E.~Toloba\altaffilmark{12,20},
L.~Van Waerbeke\altaffilmark{21},
D.~Woods\altaffilmark{21},
H.~Zhang\altaffilmark{17,18,22,23}}

\altaffiltext{1}{GEPI, Observatoire de Paris, 77 av. Denfert Rochereau, F-75014 Paris, France; E-mail:anand.raichoor@obspm.fr}
\altaffiltext{2}{Universit\'{e} Paris Denis Diderot, F-75205, Paris Cedex 13, France}
\altaffiltext{3}{Argelander-Institut f\"{u}r Astronomie, University of Bonn, Auf dem H\"{u}gel 71, D-53121 Bonn, Germany}
\altaffiltext{4}{Aix Marseille Universit\'{e}, CNRS, Laboratoire d'Astrophysique de Marseille, UMR 7326, F-13388 Marseille, France}
\altaffiltext{5}{Herzberg Institute of Astrophysics, National Research Council of Canada, Victoria, BC, V9E 2E7, Canada}
\altaffiltext{6}{Skytree, Inc., 1731 Technology Drive, Suite 700, San Jose, CA 95110, USA}
\altaffiltext{7}{INAF, Osservatorio Astronomico di Bologna, via Ranzani, 1, I-40127, Italy}
\altaffiltext{8}{Insitute of Astronomy and Astrophysics, Academia Sinica, P.O. Box 23-141, Taipei 106, Taiwan}
\altaffiltext{9}{Canada--France--Hawa\"{i}i Telescope Corporation, Kamuela, HI 96743, USA}
\altaffiltext{10}{Laboratoire AIM Paris-Saclay, CEA/IRFU/SAp, CNRS/INSU, Universit\'{e} Paris Diderot, F-91191 Gif-sur-Yvette Cedex, France}
\altaffiltext{11}{Department of Physics and Astronomy, Youngstown State University, Youngstown, OH, USA}
\altaffiltext{12}{UCO/Lick Observatory, Department of Astronomy and Astrophysics, University of California Santa Cruz, 1156 High Street, Santa Cruz, California 95064, USA}
\altaffiltext{13}{Observatoire astronomique de Strasbourg, Universit\'{e} de Strasbourg, CNRS, UMR 7550, 11 rue de l'Universit\'{e}, F-67000 Strasbourg, France}
\altaffiltext{14}{Center for Astronomy and Astrophysics, Department of Physics and Astronomy, Shanghai Jiao Tong University, 800 Dongchuan Road, Shanghai 200240, China}
\altaffiltext{15}{Shanghai Key Lab for Particle Physics and Cosmology, Shanghai Jiao Tong University, Shanghai 200240, China}
\altaffiltext{16}{Institute of Astrophysics, Pontificia Universidad Cat\'{o}lica de Chile, Av. Vicu\~{n}a Mackenna 4860, 7820436 Macul, Santiago, Chile}
\altaffiltext{17}{Department of Astronomy, Peking University, Beijing 100871, China}
\altaffiltext{18}{Kavli Institute for Astronomy and Astrophysics, Peking University, Beijing 100871, China}
\altaffiltext{19}{Department of Astronomy and Physics, St. Mary's University, Halifax, NS B3H 3C3, Canada}
\altaffiltext{20}{Observatories of the Carnegie Institution for Science, 813 Santa Barbara Street, Pasadena, CA 91101, USA}
\altaffiltext{21}{Department of Physics and Astronomy, University of British Columbia, 6224 Agricultural Road, Vancouver, B.C., V6T 1Z1, Canada}
\altaffiltext{22}{China-CONICYT Fellow}
\altaffiltext{23}{Chinese Academy of Sciences South America Center for Astronomy, Camino El Observatorio \#1515, Las Condes, Santiago, Chile}

\begin{abstract}
The Next Generation Virgo Cluster Survey is an optical imaging survey covering 104 deg$^2$ centered on the Virgo cluster.
Currently, the complete survey area has been observed in the $u^*giz$-bands and one third in the $r$-band.
We present the photometric redshift estimation for the NGVS background sources.
After a dedicated data reduction, we perform accurate photometry, with special attention to precise color measurements through point spread function-homogenization.
We then estimate the photometric redshifts with the \textit{Le Phare} and BPZ codes. We add a new prior which extends to $i_{AB} = 12.5$ mag.
When using  the $u^*griz$-bands, our photometric redshifts for $15.5 \le i \lesssim 23$ mag or $z_{\rm phot} \lesssim 1$ galaxies have  a bias $|\Delta z| < 0.02$, less than 5\% outliers, and a scatter $\sigma_{\rm outl.rej.}$ and an individual error on $z_{\rm phot}$ that increase with magnitude (from 0.02 to 0.05 and from 0.03 to 0.10, respectively).
When using the $u^*giz$-bands over the same magnitude and redshift range, the lack of the $r$-band increases the uncertainties in the $0.3 \lesssim z_{\rm phot} \lesssim 0.8$ range ($-0.05 < \Delta z < -0.02$, $\sigma_{outl.rej} \sim 0.06$, 10-15\% outliers, and $z_{\rm phot.err.} \sim 0.15$).
We also present a joint analysis of the photometric redshift accuracy as a function of redshift and magnitude.
We assess the quality of our photometric redshifts by comparison to spectroscopic samples and by verifying that the angular auto- and cross-correlation function $w(\theta)$ of the entire NGVS photometric redshift sample across redshift bins is in agreement with the expectations.
\end{abstract}



\section{Introduction}
\label{sec:intro}
 
Many fields of astronomy have entered a new era with the advent of large surveys \citep[e.g., the Sloan Digital Sky Survey; SDSS; ][]{york00}, which give access to homogeneous observations for a large number of objects (up to $10^6$-$10^8$).
Such large surveys provide invaluable information for studies of galaxy evolution and cosmology based on homogeneous measurements of a multitude of fundamental galaxy properties. 
In this context, a crucial quantity is the galaxy redshift.

While spectroscopic redshifts (spec-z's) are unambiguous measurements, they are observationally too costly for $10^6$-$10^8$ objects.
An alternative method, developed since the early '60s \citep[e.g.,][]{baum62,koo85}, is the use of photometric redshifts (photo-z's), which are estimated from photometry.
Although less precise than spec-z's, photo-z's allow consistent measurement of redshifts for large numbers of galaxies, including relatively faint ones.
The use of photo-z's for large surveys is widespread today \citep[e.g.,][]{ilbert06a,coupon09,ilbert09,bielby12,hildebrandt12,dahlen13,jouvel14} and will be essential for future missions \citep[e.g., such as \textsc{Euclid;}][]{laureijs11}.

Existing codes to estimate photo-z's can be broadly classified in two categories: template fitting and empirical estimators.
Template fitting codes (e.g.,
\textit{hyperz}: \citealt{bolzonella00};
\textsc{bpz} -- Bayesian photometric redshift: \citealt{benitez00};
\textit{Le Phare} -- Photometric Analysis for Redshift Estimate: \citealt{arnouts99,arnouts02,ilbert06a};
\textsc{Eazy} -- ÔEasy and Accurate Redshifts from Yale: \citealt{brammer08})
use empirical or theoretical galaxy spectra to find through fitting the redshift/template combination that best reproduces the observed colors, whereas empirical estimators (e.g.,
\texttt{ANN}\textit{z} -- Photometric redshifts using Artificial Neural Networks: \citealt{collister04};
\citealt{ball08};
ArborZ: \citealt{gerdes10}) use a representative sample  to train machines like neural networks and reproduce the relation between the observed colors/magnitudes and the redshifts.
The main limitations to estimate accurate photo-z's are the wavelength coverage of key spectral features (e.g., the Lyman-break and the 4,000 \AA/Balmer break), and the quality and homogeneity of the photometry.
\citet{hildebrandt10} and \citet{dahlen13} have conducted thorough analyzes on the performance of the most popular algorithms.
Both studies agree that the majority of the codes provides quantitatively similar results.
\citet{dahlen13} find that the photo-z's accuracy depends strongly on the magnitude.

We present in this paper the estimation of photo-z's for the Next Generation Virgo Cluster Survey \citep[NGVS; ][]{ferrarese12} with two template fitting codes: \textit{Le Phare} and \textsc{bpz}.
The NGVS is a comprehensive optical imaging survey of the Virgo cluster, from its core to its virial radius -- covering a total area of 104 deg$^2$ -- in the Canada-France-Hawaii Telescope (CFHT) $u^*griz$ bandpasses\footnote{The instrumental transmission curves can be found here: \url{http://www1.cadc-ccda.hia-iha.nrc-cnrc.gc.ca/megapipe/docs/filters.html} with the MegaCam instrument.
We note that the NGVS observations have been performed with the new $i$-band filter ($i$.MP9702, sometimes denoted $y$), which replaces the original $i$-band filter ($i$.MP9701) which was damaged in 2008.
Although we make the distinction in our pipeline, we write in this article $i$, regardless of the used passband, for simplicity.}.
The NGVS will serve as the optical reference survey over the Virgo cluster, and will leverage the numerous other surveys targeting Virgo at shorter and longer wavelengths, such as -- to cite only the most recent ones: 
the \textit{Galaxy Evolution Explorer (GALEX)} survey of Virgo in the ultra-violet \citep[GUViCS;][]{boselli11},
the Next Generation Virgo Cluster Survey--Infrared in the near-infrared \citep[NGVS-IR;][]{munoz14},
the Herschel Virgo Cluster Survey in the far-infrared \citep[HeViCS;][]{davies10,davies12},
or the Arecibo Legacy Fast ALFA Survey in the radio \citep[ALFALFA;][]{giovanelli05,kent08,haynes11}.

	The plan of this paper is as follows.
Section~\ref{sec:data} presents the data and their reduction.
Section~\ref{sec:photometry} details how we build the photometric catalogs.
Section~\ref{sec:zphotcalc} presents the method to estimate the photo-z's, Section~\ref{sec:zphotqual} analyzes their quality, and Section~\ref{sec:woftheta} gives a science validation of our photo-z's.
We conclude in Section~\ref{sec:conclusion}.

	In this paper, we adopt $H_0 = 70$ km s$^{-1}$ Mpc$^{-1}$,  $\Omega_m = 0.30$, and $\Omega_\Lambda = 0.70$.
All magnitudes are in the AB system and corrected for the Galactic foreground extinction using the \citet{schlegel98} maps.


\section{NGVSL\lowercase{en}S data}
\label{sec:data}

In this paper, we use a NGVS dataset whose reduction is optimized for \textit{background-science} (see below): we label this dataset NGVSLenS.
We describe in this Section the data acquisition along with their reduction.

\subsection{NGVSLenS data}

\begin{figure}[!h]
  \includegraphics[width=0.95\columnwidth]{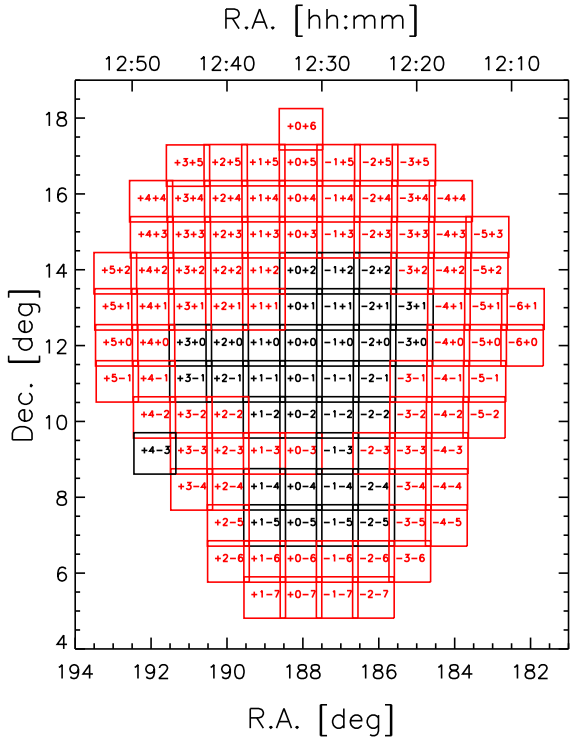}
  \caption{Layout of the NGVS survey footprint.
Pointings in black (red, respectively) have coverage in $u^*griz$ ($u^*giz$, respectively).}
\label{fig:NGVSlayout}
\end{figure}

The imaging data used in this article are from the Next Generation Virgo Cluster Survey (NGVS; P.I. L. Ferrarese).
The goals of the survey, its implementation and its observations have been described in detail in \citet{ferrarese12} and we only briefly repeat details relevant to this article.

The NGVS is a deep, 104 deg$^2$ multi-color optical imaging survey of the Virgo cluster.
All the data are obtained with the MegaCam instrument\footnote{\url{http://www.cfht.hawaii.edu/Instruments/Imaging/Megacam/}} \citep[see][]{boulade03} which is mounted on the CFHT.
MegaCam is an optical multi-chip camera with a $9\times 4$ CCD array ($2048\times 4096$ pixels in each CCD; $0\myarcsec 187$ pixel scale; $\sim$$1^{\circ}\times 1^{\circ}$ total field-of-view).
NGVS observations were carried out with 117 discrete MegaCam pointings around the \textit{NGVS central position} RA=12$^{\rm h}$32$^{\rm  m}$12$^{\rm s}$, Dec=12$^{\rm d}$00$^{\rm m}$19$^{\rm s}$, that includes Virgo's cD M87.
The exact NGVS survey layout is shown in Figure~\ref{fig:NGVSlayout}.
In this article, we follow the NGVS convention to label individual NGVS pointings  \citep[see Figure~4 of][]{ferrarese12},  which indicate the approximate separation in degrees from the NGVS central position.
For instance, pointing ``NGVS-1+2" is about one degree west and two degrees north of the NGVS center.
We however caution that the data itself are processed with a slightly different convention\footnote{We avoid the ``$+/-$" notation because of practical programming and processing reasons.} (e.g., ``NGVSm1p2" -- read ``{NGVS} minus 1 plus 2'' -- instead of ``NGVS-1+2").

The complete NGVS area (117 pointings) is covered in four SDSS-like filters: $u^*$ (CFHT identification: u.MP9301), $g$ (g.MP9401), $i$ (i.MP9702), and $z$ (z.MP9801).
Additionally, 34 pointings also benefit from $r$ (r.MP9601) band coverage (see Figure~\ref{fig:NGVSlayout}).

\begin{figure}[!h]
  \includegraphics[width=\columnwidth]{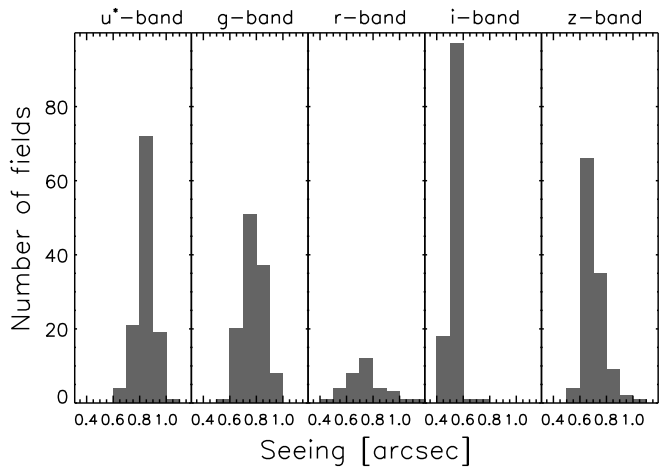}
  \caption{Seeing distributions for all co-added NGVS fields and filters.}
\label{fig:NGVSseeing}
\end{figure}

Table \ref{tab:averagequal} contains observational details and provides average quality characteristics of the final, co-added NGVS data used in this article.
It lists average observing time for the different filters, the mean limiting magnitudes and the mean seeing values with their corresponding standard deviations over all 117 NGVS pointings.
The seeing is estimated using the \textsc{SExtractor} \citep{bertin96} parameter \texttt{FWHM\_IMAGE} for stellar sources.
Our limiting magnitude, $m_{\rm lim}$, is the 5$\sigma$ detection limit in a $2\myarcsec 0$ aperture\footnote{ $m_{\rm lim}=ZP-2.5\log(5\sqrt{n_{\rm pix}} \times \sigma_{\rm sky})$, where $ZP$ is the magnitude zeropoint, $n_{\rm pix}$ is the number of pixels in a circle with radius $2\myarcsec 0$ and $\sigma_{\rm sky}$ the sky background noise variation.}.
The complete NGVS data set was obtained under very good observing conditions.
In Figure~\ref{fig:NGVSseeing} we show the full seeing distribution for all fields and filters \citep[see also Figure~8 of][]{ferrarese12}.
We specifically note the superb seeing distribution of the $i$-band: the complete survey was obtained in this filter with an exceptional seeing of $< 0\myarcsec 6$.

\begin{deluxetable}{llll}
	\tablecaption{Average characteristics of the NGVS co-added data used in this study (see the text for an explanation of the columns). \label{tab:averagequal}}
	\tablehead{
		\colhead{Filter} & \colhead{expos. time} & \colhead{$m_{\rm lim}^\dag$} & \colhead{seeing}\\
		\colhead{} & \colhead{[ks]} & \colhead{[AB mag]} & \colhead{[$''$]}
		}
	\startdata
$u^* (u.MP9301) $ & 6.3 & $25.60 \pm 0.16$ & $0.83 \pm 0.07$ \\ 
$g (g.MP9401) $ & 3.5 & $25.73 \pm 0.13$ & $0.77 \pm 0.08$ \\ 
$r (r.MP9601) $ & 2.6 & $24.68 \pm 0.50^\star$ & $0.74 \pm 0.14$ \\ 
$i (i.MP9702) $ & 2.3 & $24.41 \pm 0.13$ & $0.52 \pm 0.04$ \\ 
$z (z.MP9801) $ & 4.6 & $23.62 \pm 0.16$ & $0.70 \pm 0.08$ 
	\enddata
	\tablecomments{$^\dag$: $m_{\rm lim}$ is the 5$\sigma$ detection limit in a $2\myarcsec 0$ aperture.\\
$^\star$: for the $r$-band, the minimum and maximum values for $m_{\rm lim}$ are 23.56 and 25.52, respectively.}
\end{deluxetable}

As detailed in \citet{ferrarese12}, the NGVS data are used for a large variety of science projects.
The different applications can be split in three categories: 
(1) the \textit{foreground-science}, which will study the sources closer than the Virgo cluster;
(2) the Virgo cluster itself;
(3) and the \textit{background-science}, which uses the deep data to study the higher redshift background galaxy populations.
As outlined in \citet{ferrarese12}, the NGVS team performs different data processing and produces a large variety of data products optimized for different science applications.
\textit{Foreground}- and Virgo science, requiring dedicated data processing, will be addressed in other publications by the NGVS collaboration  \citep[e.g., ][]{durrell14}.
The current study discusses the estimation of photo-z's of background sources, which are crucial for the \textit{background-science}\footnote{though they might also be useful for Virgo cluster science for example, as in \citet{boselli11} where they permit background contamination removal.}.
Future work will include the detection of high-redshift galaxy cluster candidates (Licitra et al., \textit{in preparation}), and strong and weak lensing studies (e.g., Gavazzi et al., \textit{in preparation}).
Photo-z studies of faint background sources require high-quality, deep and carefully photometrically calibrated multi-color observations.
In the following we provide information on the preparation of the necessary data products.

\subsection{NGVSLenS data reduction}
\label{sec:data_red}

To process the NGVS data for \textit{background-science} applications, we use the algorithms and processing pipelines (\textsc{theli}) developed within the CFHTLS-Archive Research Survey \citep[CARS; see][]{erben09} and the Canada-France-Hawaii-Telescope Lensing Survey \citep[CFHTLenS; see][and \url{http://cfhtlens.org}]{heymans12,hildebrandt12,erben13}.
Both surveys originated from the Wide component of the Canada-France-Hawaii-Telescope Legacy Survey \citep[CFHTLS; ][]{gwyn12} which was also obtained with MegaCam.
In addition, the survey characteristics and the observing strategies of CFHTLS and NGVS are very similar.
This allowed for a direct transfer of our CFHTLS expertise to NGVS.
In the following we only give a very short description of our procedures to arrive at the final co-added images for NGVS.
All algorithms and prescriptions are described in detail in \citet{erben13}.
The interested reader should consult this article and the references therein.
Below, we also give a more detailed analysis of the quality from our photometric calibration which is crucial for the quality of photometric redshift estimates.

\begin{figure*}
  \fbox{\includegraphics[width=0.95\textwidth]{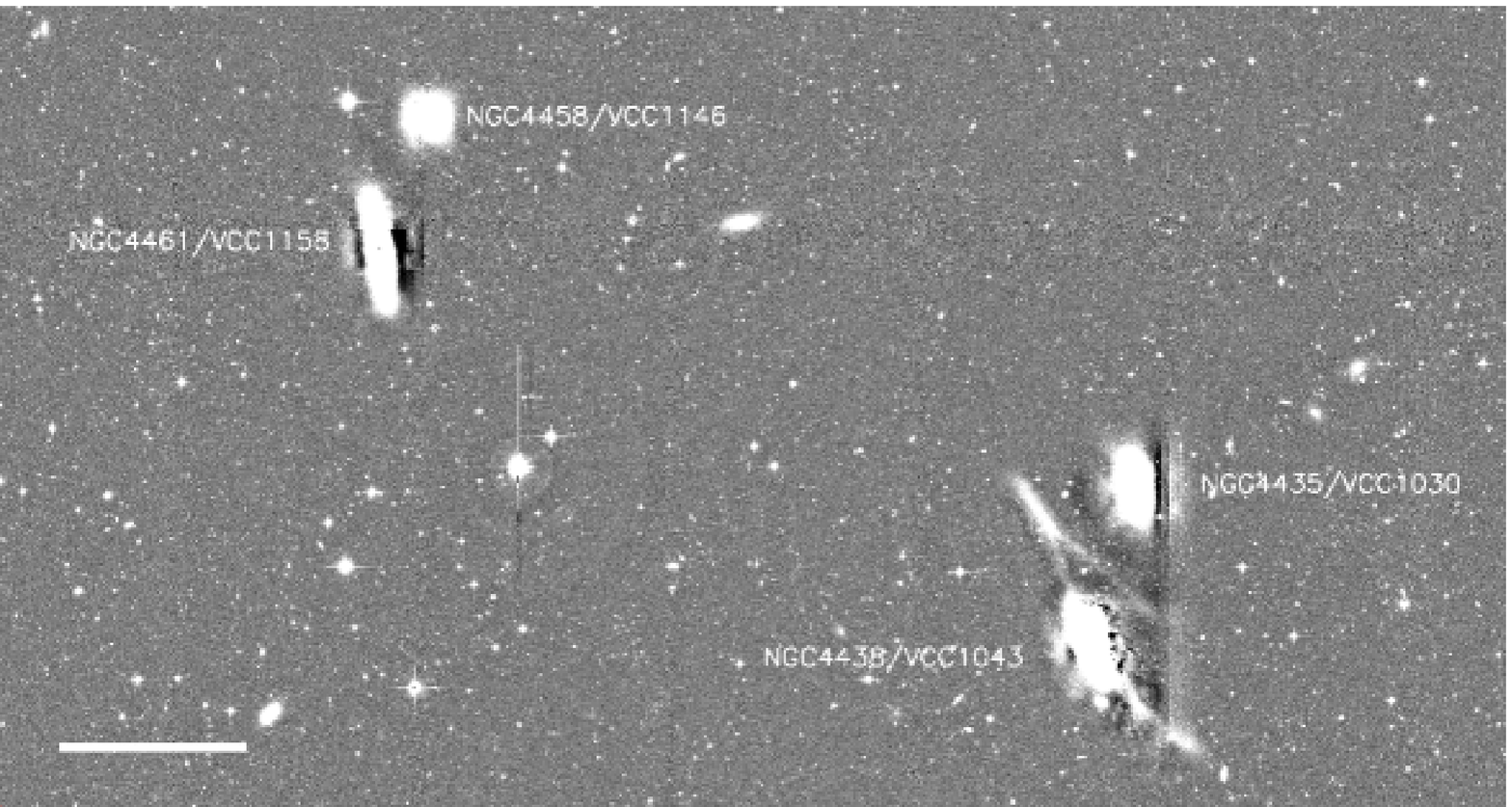}}
  \fbox{\includegraphics[width=0.95\textwidth]{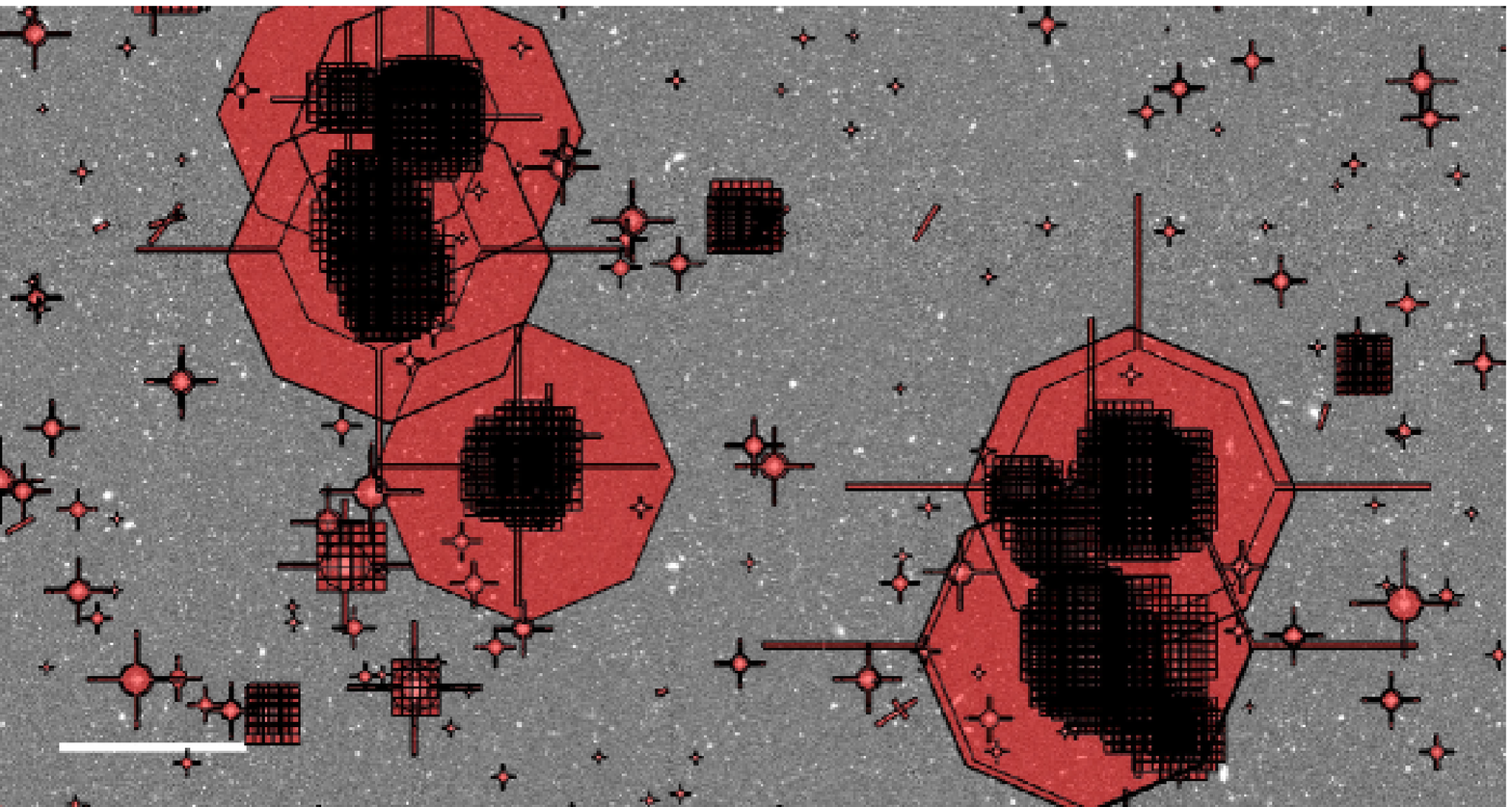}}
  \caption{Image mask of a $40\myarcmin \times 25\myarcmin$ section of the NGVS-1+1 pointing.
Our automatic masking reliably picks up bright stars and Virgo galaxies that would influence photometric analyzes of faint background galaxies through their light halos (which are an artifact of the local background subtraction).
Different mask shapes correspond to different artifacts (see text).
We identify the prominent Virgo galaxies in this field.
The ruler (white bar on the bottom-left side) is $5\myarcmin$ wide.}
\label{fig:NGVSmask}
\end{figure*}

Our NGVS data processing consisted of the following steps:
\begin{enumerate}
\item Data sample:\\
We start our data analysis with the \textsc{Elixir}\footnote{\url{http://www.cfht.hawaii.edu/Instruments/Elixir/}} preprocessed NGVS data available at the Canadian Astronomical Data Centre (CADC)\footnote{\url{http://www4.cadc-ccda.hia-iha.nrc-cnrc.gc.ca/cadc/}}.
For the current study we used NGVS observations obtained from 01/03/2008 until 12/06/2013.
The data were obtained under several CFHT programs (P.I. L. Ferrarese: 08AC16, 09AP03, 09AP04, 09BP03, 09BP04, 10AP03, 10BP03, 11AP03, 11BP03, 12AP03, 12BP03, 13AC02, 13AP03; P.I. Simona Mei: 08AF20; P.I. Jean-Charles Cuillandre: 10AD99, 12AD99 and P.I. Ying-Tung Chen: 10AT06).
From the initial set we reject all \textit{short exposed} NGVS images.
To be able to study bright cores of Virgo galaxies, the NGVS obtained, besides the primary science data, numerous \textit{short} exposures in all pointings and filters \citep[see Section 3.4 of][]{ferrarese12}.
Because these exposures would not contribute an appreciable fraction to the total exposure time in each filter, we did not consider them further.
We also do not use images whose observing conditions were marked as unfavourable by CFHT.
\item Single exposure processing:\\
The \textsc{Elixir} preprocessing includes a complete removal of the instrumental signature from raw data \citep[see also][]{magnier04}.
In addition, each exposure comes with all necessary photometric calibration information.
Therefore, we only need to perform the following processing steps on single exposures:
(1) we identify and mark individual exposure chips that should not be considered any further.
This mainly concerns chips which are completely dominated by saturated pixels from a bright star.
(2) We create sky-subtracted versions of the images.
In the context of NGVS we create a so-called \textit{local} sky-background subtraction optimal for the study of faint background galaxies \citep[see also Figure~11 of][]{ferrarese12}.
In addition, we create a weight image for each science chip.
It gives information on the relative noise properties of individual pixels and assigns a weight of zero to defective pixels (such as cosmic rays, hot and cold pixels, areas of satellite tracks).
(3) We use \textsc{SExtractor} to extract high $S/N$ sources\footnote{We consider all sources having at least 5 pixels with at least $5\sigma$ above the sky-background variation.} from the science image and weight information.
These source catalogs are used to astrometrically and photometrically calibrate the data in the next processing step.
In addition we perform an analysis of the PSF anisotropy and use this information to reject images showing high stellar ellipticities. 
Highly elongated point sources are a good indication of tracking issues or other severe problems during an exposure.
\item Astrometric and Photometric calibration: \\
We use the \textsc{scamp} software\footnote{\url{http://www.astromatic.net/software/scamp}} \citep[see][]{bertin06} to astrometrically calibrate the NGVS survey.
We use the Two Micron All Sky Survey \citep[2MASS;][]{skrutskie06} as astrometric reference and calibrate separately each filter from the NGVS patch (i.e.,  all fields) with \textsc{scamp}.
Once an astrometric solution is established we use overlap sources from individual exposures to establish an internal, relative photometric solution for all exposures.
We reject all exposures with an absorption of more than 0.2 magnitudes\footnote{More than $95\%$ of the NGVS data has been obtained under at least good photometric conditions with an absorption of 0.05 magnitudes or less. Our rejection limit of 0.2 magnitudes turned out to be a very good conservative limit to reject the small fraction of images observed under poor photometric conditions.} and rerun \textsc{scamp} on the remaining images.
With the relative photometric solution and the \textsc{Elixir} zeropoint information we estimate a patch-wide photometric zeropoint for each filter.
\item Image co-addition and mask creation: \\
The next step of our image processing co-adds the sky-subtracted exposures belonging to a pointing/filter combination with the \textsc{swarp} program\footnote{\url{http://www.astromatic.net/software/swarp}} \citep[see][]{bertin02}.
The stacking is performed with a statistically optimally weighted mean which takes into account sky-background noise, weight maps and the \textsc{scamp} relative photometric zeropoint information.
As a final step we use the \textsc{automask} tool \citep[see][]{dietrich07} to create image masks for all pointings.
These masks flag bright, saturated stars and areas which would influence the analysis of faint background sources.
For the NGVS, a reliable masking of bright Virgo members is particularly important for our purposes.
The 117 generated masks have been visually checked (A.R.).
In a typical NGVS pointing we loose about 20\%-30\% of the area because of masking, as can be seen in Figure~\ref{fig:NGVSmask}, where we display those masks for a section of a field.
In the figure we note that several algorithms and template masks are used to mask different \textit{artifacts}, e.g., bright stars, short asteroid trails or large-scale bright objects.
The artifacts we consider and the way we treat them is described in more detail in \citet{erben09}.
\item Photometric calibration tied to the SDSS: \\
The final step is specific to the NGVS, i.e.,  it has not been implemented in the released CFHTLenS.
Taking advantage of accurate internal photometric stability of the SDSS, and of its full coverage of the NGVS field, we tie our photometric calibration to the SDSS.
We retrieve clean stars from the SDSS-DR10 \citep{ahn14} and convert their Petrosian magnitude to the MegaCam photometric system using Equation (4) of \citet{ferrarese12}.
For each field and each filter available, we then measure the \texttt{MAG\_AUTO} with \textsc{SExtractor} and correct for the offset between the two catalogs by taking the median value for a subset of bright non-saturated stars ($\sim$500 per field).
The typical uncertainty in this calibration step is of 0.05 mag in the $griz$-bands and of 0.10 mag in the $u$-band.
We note that the \textsc{theli} photometry is homogeneous over the NGVS field, with a field-to-field standard deviation of $\sim$0.03 mag, as found in \citet{erben13}.
\end{enumerate}

\section{Photometric catalogs}
\label{sec:photometry}

For simplicity, we hereafter are referring to NGVSLenS when referring to NGVS or the NGVSLenS dataset, and similarly to CFHTLenS with CFHTLS or the CFHTLenS dataset.

In this Section we describe the procedure to build the photometric catalogs when the five $u^*griz$-bands are available (the procedure is similar when only the four $u^*giz$-bands are available).
As studied in detailed in \citet{hildebrandt12}, a requirement to estimate precise photo-z's is accurate photometry, in particular high precision color measurements.
To do so, we implement the following procedure \citep[\textit{global} mode of][]{hildebrandt12}:
the $i$-band, which has the best seeing ($0.52\arcsec \pm 0.04 \arcsec$), is used to detect objects and estimate their total magnitude;
then, for each field, all images are first homogenized to the same PSF and then used to estimate accurate colors.

\subsection{Global PSF homogenization}
The photometric catalogs are constructed as described in \citet{hildebrandt12}, which also describe the \textit{global} PSF homogenization that is necessary to measure unbiased colors.
According to the \citet{hildebrandt12} analysis, done on the CFHTLenS data having properties similar to the NGVSLenS data ones (see Section \ref{sec:zphotqual} and Appendix \ref{sec:lensprop}), the quality of the photo-z's obtained assuming a constant PSF across each field (\textit{global} mode) provides satisfactory results, even when compared to that obtained when accounting for the PSF variations across each field (\textit{local} mode).
For this analysis, we hence consider that it is satisfactory to make the approximation that the PSF is constant over each field (1 deg$^2$) and can be described by a single Gaussian with width $\sigma_{\rm PSF}$.

For each field, we identify the band which has the largest seeing ($\sigma_{\rm PSF,worst}$) and we bring the other four images to the same seeing by convolving them with a two-dimensional Gaussian filter.
For instance if the $X$-band image has a PSF width $\sigma_{\rm PSF,X}$, we convolve it with a Gaussian filter of width $\sqrt{\sigma_{\rm PSF,worst}^2-\sigma_{\textrm{PSF},X}^2}$.
The 117 values values of $\sigma_{\rm PSF,worst}$ have a mean value of $0.85 \pm 0.07$, and the band with $\sigma_{\rm PSF,worst}$ is the $u^*$- ($g$-, $r$-, $i$-, and $z$-, respectively) band in 58 (45, 7, 0, and 7, respectively) fields.

\subsection{Photometry method}
\label{sec:phot_method}
Multicolor catalogs in the $u^*griz$ bands are extracted from a set of these PSF homogenized images using \textsc{SExtractor} in dual-image mode, using the un-convolved $i$-band image as the detection image.
One \textsc{SExtractor} run is performed on the un-convolved $i$-band image for detection, structural measurements, and estimation of the total $i$-band magnitudes (\textsc{SExtractor} \texttt{MAG\_AUTO}).
Five \textsc{SExtractor} runs (dual-image mode, with the un-convolved $i$-band image as the detection image) are then performed with the PSF-matched $u^*griz$ images to measure accurate colors.
Indeed, the obtained isophotal magnitudes \texttt{MAG\_ISO} are measured on the same physical apertures, as we are using the same pixels (dual-mode) of PSF-matched images.
From this procedure, we obtain accurate measurements for the colors and the $i$-band total magnitude; estimation of the total magnitude in the $u^*grz$ bands can be obtained with \citep{hildebrandt12}:
\begin{equation}
	X_{\rm tot} = i_{\rm tot} + (X-i), \; \textnormal{with} \; X \in \{u^*,g,r,z\},
	\label{eq:magtot}
\end{equation}
where $X_{\rm tot}$ and $i_{\rm tot}$ are the total magnitudes in the $X$ and $i$ bands respectively, and $(X-i)$ is the corresponding color index.
Note that we do not use the estimations of the total magnitude in the $u^*grz$ bands.
Also, we caution that these estimations will only be valid if the galaxies do not present strong color gradients.

\subsection{Photometric errors}
\begin{figure}[!h]
	\includegraphics[width=\columnwidth]{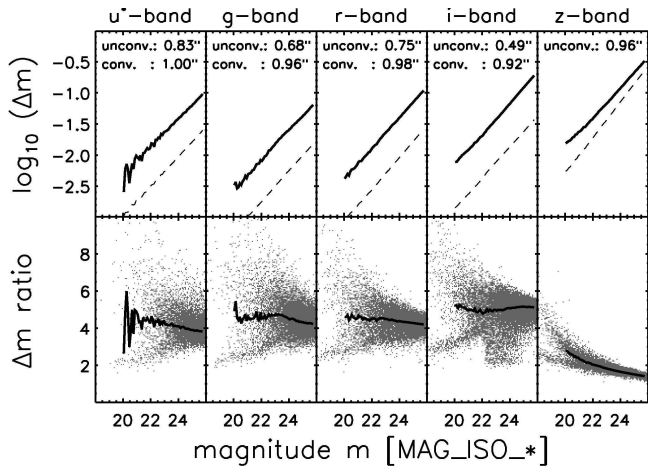}
	\caption{Re-estimated magnitude error ($\Delta m$) properties for the NGVS+0+0 field.
The continuous and dashed lines show the median values for $20 \le m \le 26$ mag.
\textit{Top panel}: we show $\Delta m$ as a function of magnitude $m$; our re-estimated errors are shown as continuous lines while \textsc{SExtractor} errors are denoted by dashed lines.
We report for each band the seeing before and after convolution (for the NGVS+0+0 field, the $z$-band image is the one with the worst seeing, hence it is not convolved).
\textit{Bottom panel}: we show the ratio of our $\Delta m$ to \textsc{SExtractor}'s $\Delta m$ as a function of magnitude $m$;  points illustrate the individual distribution of objects (for clarity, we plot only 1 out of 5 points).
\textsc{SExtractor} errors are on average underestimated by a factor of $\sim$4-5 on convolved images and of $\sim$2 on un-convolved images.}
\label{fig:photrms}
\end{figure}

In this study, we pay special attention to the photometric error estimation.
Noise correlation introduced by image resampling during the reduction artificially decreases the pixel-to-pixel rms variations $\sigma_1$, which leads to an underestimation of the flux errors estimated by \textsc{SExtractor} \citep[e.g.][]{casertano00}.
This effect is known and the flux error underestimation factor for the CFHT/MegaCam optical bands is of the order of 1.5 \citep[e.g.,][]{ilbert06a,coupon09,raichoor12}.
However, it can be much larger on convolved images and the flux error underestimation factor can be of the order of 5, as shown by our data analysis (see Figure \ref{fig:photrms}).
This phenomenon can be more pronounced in the presence of fringing \citep{ferrarese12}.
We choose the following method to estimate the true background fluctuations, $\sigma_{\rm bkg}$: regardless of whether the measurement is performed on a convolved or un-convolved image, we estimate $\sigma_{\rm bkg}$ in the un-convolved image, as it is not affected by the convolution process\footnote{Indeed, the $\sigma_1$ in the convolved image will have an artificially low value due to the noise correlation introduced by the convolution process.}.

For each of our 502 un-convolved images, we estimate $\sigma_{\rm bkg}$ by placing 2,000 random apertures of a given size, which do not overlap with any detected objects \citep[e.g.][]{labbe03, gawiser06}.
We use circular apertures of area $n_{\rm pix}$ centered at integer pixels and describe them by the linear size of the aperture defined as $N = \sqrt{n_{\rm pix}}$, and fit a Gaussian to the histogram of aperture fluxes to yield $\sigma_{\rm bkg}(N)$, the  background fluctuation for a given aperture $N$.
We apply this method for $0\arcsec < N < 3\arcsec$.
Then we fit the obtained $\sigma_{\rm bkg}(N)$ curve as a function of $N$ following \citet{labbe03} formalism:
\begin{equation}
	\sigma_{\rm bkg}(N) = \sigma_1 \times (aN + bN^2).
	\label{eq:rmsfit}
\end{equation}
$\sigma_1=\sigma(1)$ is the pixel-to-pixel rms variations, measured through 1 pixel apertures for each un-convolved image for a given band and a given field.
We present in Table \ref{tab:bkgfit} the median of the 502 fitted values for $\sigma_1$, $a$, and $b$.

\begin{deluxetable}{llll}
	\tablecaption{Median values of the image noise parameters fitted with Eq.(\ref{eq:rmsfit}). \label{tab:bkgfit}}
	\tablehead{
		\colhead{Filter} & \colhead{$\sigma_1$} & \colhead{$a$} & \colhead{$b$}\\
		\colhead{} & \colhead{[count.s$^{-1}$]} & \colhead{} & \colhead{}
		}
	\startdata
$u$ & $0.008\pm0.001$ & $0.995\pm0.037$  & $0.046\pm0.005$\\
$g$ & $0.020\pm0.002$ & $1.029\pm0.038$  & $0.046\pm0.007$\\
$r$ & $0.036\pm0.022$ & $1.054\pm0.047$  & $0.041\pm0.012$\\
$y$ & $0.040\pm0.005$ & $1.025\pm0.039$  & $0.044\pm0.007$\\
$z$ & $0.030\pm0.004$ & $0.865\pm0.059$  & $0.110\pm0.028$
	\enddata
\end{deluxetable}

Using error propagation and Poissonian uncertainties, the magnitude uncertainty $\Delta m$ for an object with a measured flux $F$ (in ADU.s$^{-1}$) and a pixel area $n_{\rm pix}$ is obtained by accounting for the background noise and the Poissonian noise intrinsic to the object:
\begin{equation}
\Delta m = \frac{2.5}{ln 10} \times \frac{1}{w} \times \frac{\sqrt{F/g + [\sigma_{\rm bkg}(\sqrt{n_{\rm pix}})]^2}}{F},
	\label{eq:magerr}
\end{equation}
where $g$ is the gain and $w$ is the square root of the number of single frames that contribute to the considered pixels divided by the number of single frames used to build the un-convolved image \citep[which gives an estimation of the weight on the considered pixels; cf.][]{erben13}.
We use the \textsc{SExtractor} outputs\footnote{For \texttt{AUTO} (isophotal, respectively) magnitudes, the flux $F$ is given by \texttt{FLUX\_AUTO} (\texttt{FLUX\_ISO}, respectively) and the area $n_{\rm pix}$ is given by  $\pi$$\times$\texttt{A\_IMAGE}$\times$\texttt{B\_IMAGE}$\times$\texttt{KRON\_RADIUS}$^2$ (\texttt{ISOAREAF\_IMAGE}, respectively).} for the flux $F$ and the pixel area $n_{\rm pix}$ estimations.

We illustrate in Figure~\ref{fig:photrms} how our estimated photometric uncertainties compare with those of \textsc{SExtractor} for the NGVS+0+0 field.
When the images are not convolved ($z$-band for this field), we recover the usual \textsc{SExtractor} underestimation of $\sim$1.5 because of pixel correlation.
However, we see that the underestimation is much greater when \textsc{SExtractor} is run on a convolved image, and that this underestimation is a function of the convolution kernel and of the object's magnitude (and size).
When plotting the individual objects (bottom panels), we remark that stars have a different behavior, because of their small $n_{\rm pix}$, compared to galaxies of similar magnitude.\\

\section{Photometric redshifts estimation}
\label{sec:zphotcalc}

With the photometric catalogs described in the previous section in hand, we are able to estimate the photo-z's.
We describe in this section the procedure used to estimate them.
In Table \ref{tab:photoz_setup}, we summarize the setup parameters used in this analysis.

\begin{deluxetable*}{ll}
	\tablewidth{0pt}
	\tablecaption{Photo-z setup parameters summary. \label{tab:photoz_setup}}
	\tablehead{
		\colhead{Parameter} & \colhead{Comment}
		}
	\startdata
Template set			&	El, Sbc, Scd, Im, SB2, SB3 \citep{capak04}\\
Prior					&	Appendix~\ref{sec:prior} (\textit{Le Phare} prior for $i>20$ mag, extended down to $i=12.5$ mag)\\
$E(B-V)$ reddening		&	\textit{Le Phare}: $0 \le E(B-V) \le 0.25$; \textsc{BPZ}: none\\
Reddening law			&	\textit{Le Phare}: \citet{prevot84}; \textsc{BPZ}: none\\
Minimum photometric error&	$rgiz$-band: 0.05 mag; $u$-band: 0.10 mag
	\enddata
\end{deluxetable*}

\subsection{Code: \textit{Le Phare} and \textsc{BPZ}}
In the present study, we use two template-based codes to estimate photo-z's:
\textit{Le Phare}\footnote{\href{http://www.cfht.hawaii.edu/\~arnouts/LEPHARE/lephare.html}{http://www.cfht.hawaii.edu/$\sim$arnouts/LEPHARE/lephare.html}} \citep{arnouts99,arnouts02,ilbert06a}
and
\textsc{bpz} \citep{benitez00,benitez04,coe06}.
In addition to having been widely used and tested, \citet{hildebrandt10} and \citet{dahlen13} have shown that these two codes provide satisfactory results.

\subsection{Templates}
\label{sec:templates}
For both codes, we use the recalibrated template set of \citet{capak04}, which is built from the four \citet{coleman80} observed galaxy spectra (El, Sbc, Scd, Im), with two additional observed starburst templates from \citet{kinney96}.
We note that, when running \textit{Le Phare} and \textsc{bpz}, those six templates are linearly interpolated into $\sim$60 templates, so to have a better sampling of the color space.

\begin{figure}[!h]
	\includegraphics[width=\columnwidth]{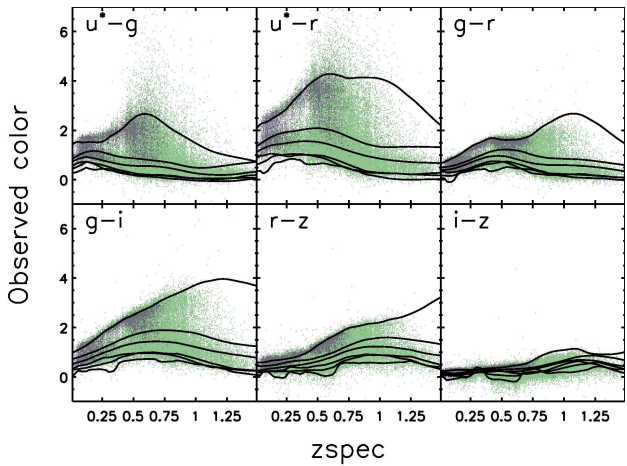}
	\caption{Observed colors as a function of $z_{\rm spec}$: magenta and green dots show our NGVSLenS and CFHTLenS spectroscopic samples, respectively (both described in Section\ref{sec:samples}); black lines represent the colors predicted by our adopted template set.
Our template set satisfactorily reproduces the observed colors.}
	\label{fig:template_color}
\end{figure}

A requirement of our template set is its ability to reproduce the observed colors.
In Figure~\ref{fig:template_color}, we display the observed colors for our spectroscopic sample ($\sim$83,000 galaxies; described in Section~\ref{sec:samples}), along with the colors predicted by the templates.
Our templates cover in a satisfactory way the observed colors.
We note that galaxies having a $u-r$ color redder than the models are a minority: for instance, less than 3\% of galaxies with $0.4 < z_{\rm spec} < 0.9$ have $u-r>5$ mag.\\

\textit{Le Phare} offers the possibility to include galaxy internal reddening, $E(B-V)$ as a free parameter during the fit.
When running \textit{Le Phare} for spectral types later than Sbc, we let $E(B-V)$ as a free parameter ($0 \le E(B-V) \le 0.25$) using the Small Magellanic Cloud (SMC) extinction law \citep{prevot84} \citep[see][]{coupon09}.
We have tested that our results of Section~\ref{sec:zphotqual} are independent of the use of an other extinction law \citep[Large Magellanic Cloud (LMC),][]{fitzpatrick86} or a larger range of possible reddening ($0 \le E(B-V) \le 0.5$).

\subsection{Fitting procedure and new prior}
Overall, both codes run in a similar way using a Bayesian approach.
This approach, and more precisely the use of a prior, allows the estimate of robust photo-z's; its advantage over the maximum-likelihood method is described in detail in \citet{benitez00}.
Briefly, in the maximum-likelihood method, \textit{a priori} assumptions are implicitly made on the choice of the explored parameter space.
The Bayesian approach, with the use of a prior, implements \textit{a priori} knowledge in a more systematic and detailed way.
Regarding the photo-z estimation, thanks to the existence of large, intensive spectroscopic surveys  coupled with imaging (e.g., SDSS; zCOSMOS: \citealt{lilly07}; DEEP2: \citealt{cooper08}; VVDS: \citealt{le-fevre13}) we have some precise and statistically robust knowledge about the relationship between spec-z, magnitude, and spectral type.
This \textit{a priori} knowledge is used to favor more physical solutions and brings essential constraints when there are few bandpasses to estimate the photo-z's, as it is the case in our work.

We now succinctly present the Bayesian approach \cite[please refer to][for a detailed presentation]{benitez00}.
For a given galaxy with an observed magnitude $m_0$ in a reference band ($i$-band here) and observed colors $C$, the posterior $p(z | m_0,C)$, i.e.,  the probability for this galaxy to be at redshift $z$ given the observed data, can be expressed as a sum of probabilities over the basis formed by the different spectral distribution types, $T$, belonging to our template set (see Section \ref{sec:templates} and Table \ref{tab:photoz_setup}):
\begin{equation}
	p(z | m_0, C) =  \sum_{T} p(z, T | m_0,C),
	\label{eq:bayes1}
\end{equation}
where $p(z, T | m_0,C)$ is the probability of the galaxy redshift being $z$ and the galaxy spectral template type being $T$.
According to Bayes' theorem,  $p(z, T | m_0,C)$ is proportional to the product of the likelihood $p(C | T, z)$ of observing those colors for a galaxy of spectral template type $T$ at redshift $z$ and of the prior $p(z,T | m_0)$, which translates the \textit{a priori} probability for a galaxy of magnitude $m_0$ to be at redshift $z$ and have a spectral template type $T$:
\begin{equation}
	p(z, T | m_0,C) \propto p(C | T, z) \times p(z , T | m_0).
	\label{eq:bayes2}
\end{equation}

To account for zeropoint uncertainty (see Section \ref{sec:data_red}), we add in quadrature an uncertainty of 0.05 mag in the $griz$-bands and of 0.10 mag in the $u$-band.
Those account for the typical uncertainties in our photometric calibration explained in Section \ref{sec:data_red}.
Both codes provide the redshift posterior distribution.
We take as the photo-z estimate the median of this posterior distribution.
We choose to use the median of the p(z) because it improves the redshift estimation in the most difficult cases in which the algorithm cannot define a clear peak of the distribution (see also the \citealt{dahlen13} analysis when comparing results from different photo-z algorithms).
This corresponds to the \texttt{Z\_ML} output of \textit{Le Phare}; \textsc{bpz} does not provide this output: we compute it based on the output posterior.
Regarding the photo-z uncertainties, we use the boundary of the interval including 68\% of the redshift probability distribution function.
\textit{Le Phare} outputs those as \texttt{Z\_ML68\_LOW} and \texttt{Z\_ML68\_HIGH};
\textsc{bpz} provides only 95\% uncertainty on $z_{\rm phot}$: for each object, we compute, based on the output posterior, the 68\% confidence interval as defined in \textit{Le Phare}.

\subsubsection{Introduction of a new prior}
\textit{Le Phare} and \textsc{bpz} were designed for high redshift studies: both codes use similar priors for $i>20$ mag galaxies, built with observed data.
However, the priors used for $i<20$ mag galaxies are not calibrated on observed data and, as a consequence, do not provide satisfying  constraints.
A direct outcome is low-quality photo-z's for $z_{\rm spec} \lesssim 0.2$ objects, for which the photo-z's have either a large scatter with \textit{Le Phare} (e.g., see Figure~11 of the CFHTLS/Wide T0007 paper\footnote{\url{ftp://ftpix.iap.fr/pub/CFHTLS-zphot-T0007/cfhtls\_wide\_T007\_v1.2\_Oct2012.pdf}}) or are biased towards high values with \textsc{bpz} \citep[e.g., see Figure~4 of][]{erben09}.
As a result of the large area covered by the NGVSLenS, $i<20$ mag galaxies represent a non-negligible fraction of our sample: \citet{hildebrandt12} already noticed this issue with the CFHTLenS data -- the \textsc{bpz} prior biasing the posterior against low photo-z's -- and implemented an \textit{ad hoc} solution.
We tackle this issue in a more systematic way by extending the prior to bright objects.
We use the SDSS Galaxy Main Sample spectroscopic survey \citep{york00, strauss02, ahn14} to establish the prior for $13 < i \le 17$ mag galaxies, and extrapolate the prior for $17 < i < 20$ mag galaxies.
The construction of this extended prior is detailed in Appendix~\ref{sec:prior}.

\subsubsection{No photometric re-calibration}
When using the template fitting method, it is common to add some photometric offsets during the fitting \citep[e.g.,][]{brodwin06,ilbert06a, hildebrandt10,dahlen13} because they can improve the accuracy of the estimated photo-z's.
They may correct for various effects, such as imprecise photometric calibration, mismatch between the used templates and spectral energy distributions of observed galaxies, imprecise filter throughputs, or different properties of source images when using multi-color catalogs.
These offsets are calculated with an iterative process comparing the colors predicted from the templates with the  colors measured for a spectroscopic subsample.

We have tested the computation of such offsets (using the bright objects of our spectroscopic sample described in Section~\ref{sec:samples}).
We find small offsets ($<0.03$ mag in absolute value) and hence do not use them in the present study.
Our approach is in agreement with the analysis of \citet{hildebrandt12}, who concluded that using PSF-matched photometry decreases the offset amplitude.

\section{Photometric redshift accuracy}
\label{sec:zphotqual}

We present in this Section an analysis of our photo-z's to quantify their accuracy.

As detailed below, our spectroscopic sample over the NGVSLenS field is rather shallow ($z \lesssim 0.8$) and highly biased at $z \gtrsim 0.3$.
In order to assess the quality of our photometric redshifts up to $z < 1.5$, we use the CFHTLenS data, which are covered by deep and intensive spectroscopic surveys.

Those CFHTLenS data have been imaged with the same telescope, instrument, filters (except for the $i$-band filter which was replaced), have similar depth, and have been reduced with the same \textsc{theli} pipeline.
Starting from the CFHTLenS \textsc{theli} coadded-images, we re-estimate for the CFHTLenS the photometry and photo-z's, with the \textsc{theli} pipeline including our modifications described in the previous sections (including the photometric calibration tied to the SDSS; see (v) of Section~\ref{sec:data_red}).
We confirm \textit{a posteriori} the close similarity of the CFHTLenS and NGVSLenS datasets in  Appendix~\ref{sec:lensprop}.

We can thus compare our photo-z's with two complementary spectroscopic samples, over the NGVSLenS and the CFHTLenS fields, as described in the next section.

In this analysis, we exclude very low redshift ($z \le 0.01$) objects, mainly Virgo objects, as those are either spectroscopically confirmed (Virgo galaxies) or have redshifts difficult to constrain with optical data only (Virgo globular clusters -- GCs -- and Ultra-compact dwarf galaxies -- UCDs).
\citet{munoz14} show that near-infrared data are crucial to diagnose those populations.
Those objects, along with Galactic stars, are excluded either using the spec-z for the spectroscopic samples, either using the criteria presented in Appendix~\ref{sec:StarGCrem} for the photometric samples.
We remark that our pipeline computes a photo-z for those objects, but we do not analyze it here.

In Section~\ref{sec:samples}, we present the samples that we use to analyze our photo-z's.
We quantify, as a function of magnitude or redshift, the accuracy of our photo-z's when they are estimated with the $u^*griz$-bands (Sections \ref{sec:zphot_zspec_ugriz} and \ref{sec:zphoterr}) or with the $u^*giz$-bands (Section~\ref{sec:zphot_nor}).
Section~\ref{sec:joint} presents a joint analysis of photo-z dependence on magnitude and redshift.

\subsection{Samples definition}
\label{sec:samples}

In this section, we present and define the samples used to analyze the accuracy of our photo-z's.
We first present the two spectroscopic samples covering the NGVSLenS and the CFHTLenS.
We then present the photometric samples covering the NGVSLenS.
The properties of those samples are summarized in Table \ref{tab:samples}.

\begin{deluxetable*}{lllllll}
	\tablewidth{0pt}
	\tablecaption{Properties of the samples used in Section~\ref{sec:zphotqual}. \label{tab:samples}}
	\tablehead{
		\colhead{Survey} & \colhead{Number of galaxies} & \colhead{Area} & \colhead{Band coverage} &
			\colhead{$\langle z_{\rm spec} \rangle^\dag$} & \colhead{$\langle i \rangle^\dag$} & \colhead{Reference}\\
		\colhead{} & \colhead{[$10^3$]} & \colhead{[deg$^2$]} & \colhead{} & \colhead{} & \colhead{[mag]}	 & \colhead{}
		}
	\startdata
\multicolumn{7}{c}{Spectroscopic sample over the NGVSLenS field ($0.01 \le z_{\rm spec} < 1.5$)}\\
\hline
SDSS/GalMS		&	9.2		&	104		&	$u^*griz/u^*giz$	&	$0.11\pm0.05$	&	$16.8\pm1.1$	& (1)\\
SDSS/notGalMS	&	14.3		&	104		&	$u^*griz/u^*giz$	&	$0.48\pm0.13$	&	$19.3\pm0.8$	& (2)\\
AAT				&	1.4		&	$\sim$30	&	$u^*griz/u^*giz$	&	$0.15\pm0.09$	&	$18.2\pm0.6$	& (3)\\
MMT				&	1.1		&	$\sim$4	&	$u^*griz$			&	$0.19\pm0.10$	&	$18.7\pm0.8$	& (4)\\
Keck				&	0.1		&	-		&	$u^*griz/u^*giz$	&	$0.72\pm0.37$	&	$23.2\pm1.2$	& (5)\\
\textbf{Compiled}	&	\textbf{26.1}&	\textbf{104}&	\textbf{$u^*griz/u^*giz$}	&	\textbf{0.32}$\pm$\textbf{0.21}	&	\textbf{18.3}$\pm$\textbf{1.5}	& -\\
\hline
\multicolumn{7}{c}{Spectroscopic sample over the CFHTLenS field ($0.01 \le z_{\rm spec} < 1.5$)}\\
\hline
SDSS/GalMS		&	0.8		&	$\sim$24	&	$u^*griz$			&	$0.11\pm0.05$	&	$16.6\pm0.6$	& (1)\\
SDSS/notGalMS	&	5.6		&	$\sim$42	&	$u^*griz$			&	$0.41\pm0.21$	&	$19.0\pm1.6$	& (2)\\
VVDS/F22		&	4.0		&	$\sim$3	&	$u^*griz$			&	$0.52\pm0.23$	&	$21.4\pm0.9$	& (6)\\
VVDS/F02		&	5.0		&	$<1$		&	$u^*griz$			&	$0.69\pm0.30$	&	$22.7\pm1.2$	& (6)\\
DEEP2/EGS		&	12.1		&	$<1$		&	$u^*griz$			&	$0.71\pm0.32$	&	$22.5\pm1.2$	& (7)\\
VIPERS			&	29.6		&	$\sim$17	&	$u^*griz$			&	$0.68\pm0.16$	&	$21.7\pm0.8$	& (8)\\
\textbf{Compiled}	&	\textbf{57.2}&	\textbf{42}	&	\textbf{$u^*griz$}	&	\textbf{0.65}$\pm$\textbf{0.25}	&	\textbf{21.6}$\pm$\textbf{1.5}	& -\\
\hline
\multicolumn{7}{c}{Photometric samples over the NGVSLenS field}\\
\hline
NGVSLenS/phot23	&	576.7	&	$\sim$30	&	$u^*griz$			&	-	&	$21.9\pm 1.0$	& This paper\\
NGVSLenS/phot24	&	1,263.5	&	$\sim$30	&	$u^*griz$			&	-	&	$22.8\pm 1.1$	& This paper\\
\hline
	\enddata
	\tablecomments{$^\dag$: mean and standard deviation.}
	\tablerefs{
		(1): \citet{strauss02};
		(2): \citet{eisenstein01,dawson13};
		(3): Zhang et al. (\textit{submitted}); Zhang et al. (\textit{in preparation});
		(4): Peng et al. (\textit{in preparation});
		(5): Guhathakurta et al. (\textit{in preparation});
		(6): \citet{le-fevre05,le-fevre13}
		(7): \citet{davis03,newman13a}
		(8): \citet{guzzo13}.
}
\end{deluxetable*}

\subsubsection{Spectroscopic sample over the NGVSLenS}
\begin{figure*}[!ht]
	\begin{tabular}{l @{\hskip 50pt} r}
		\includegraphics[width=0.90\columnwidth]{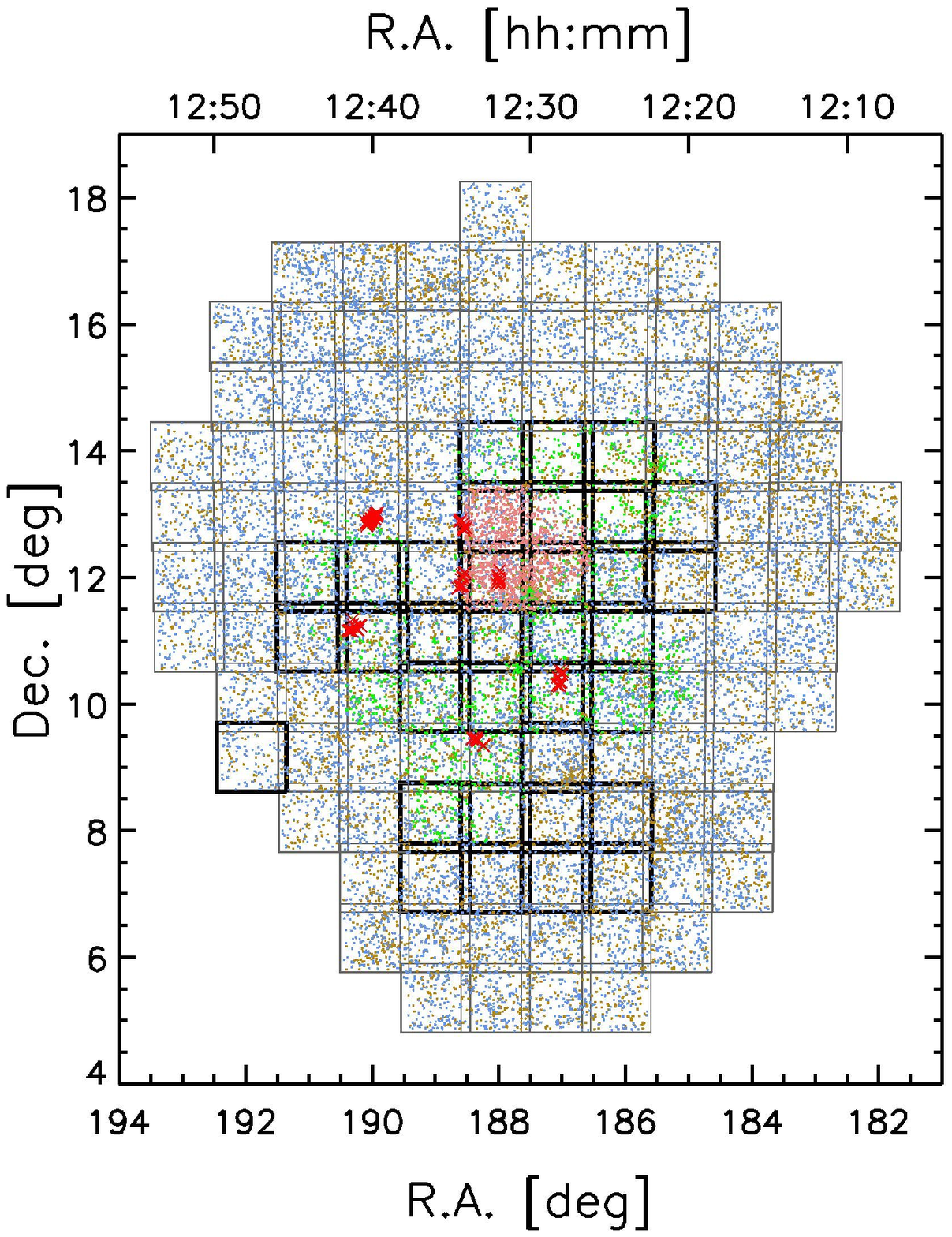} &
		\includegraphics[width=0.90\columnwidth]{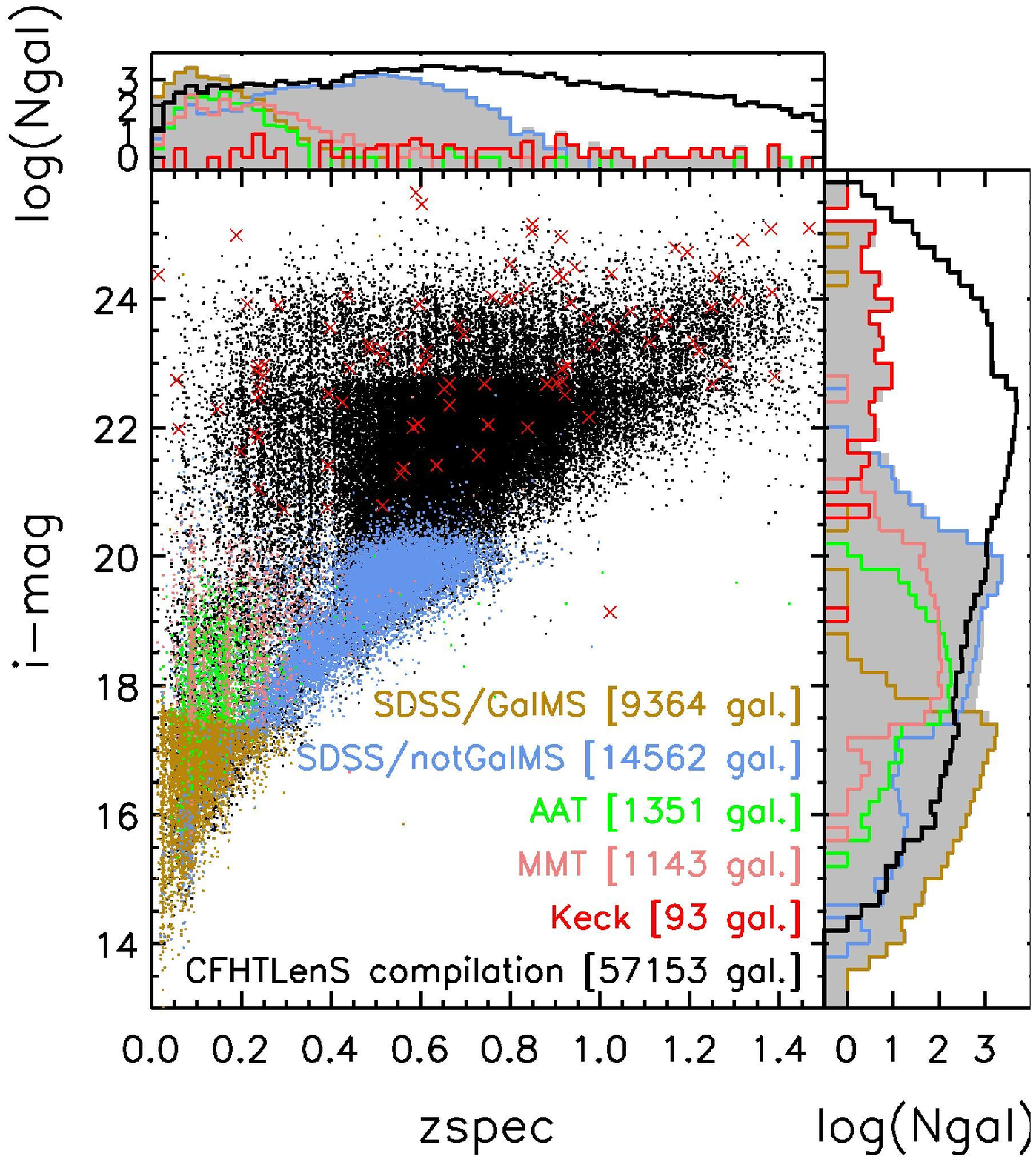}\\
	\end{tabular}
	\caption{Spectroscopic samples properties.
Spec-z's from the SDSS/Galaxy Main Sample (SDSS/other programs, AAT, MMT, Keck, respectively) survey are in gold points (blue points, green points, coral points, red crosses, respectively) symbols.
The black points are the CFHTLenS spectroscopic sample (SDSS; VVDS F02/F22; DEEP2/EGS; VIPERS W1/W4).
\textit{Left panel}: we show the spatial distribution of the NGVSLenS spectroscopic sample; the fields covered by the $u^*griz$ bands have thick black outlines.
\textit{Right panel}: we show $i$-band magnitude as a function of $z_{\rm spec}$; the filled grey histograms represent the whole NGVSLenS spectroscopic sample.}
\label{fig:zspecsample}
\end{figure*}

The NGVSLenS field is covered by several spectroscopic surveys having different target selections.
The entire NGVSLenS field is covered by the SDSS, providing $\sim$23,500 galaxy spec-z's: $\sim$40\% come from the Galaxy Main Sample, which is magnitude limited \citep[$r\le17.77$;][]{strauss02} and have $\langle z_{\rm spec} \rangle = 0.11 \pm 0.05$; the remaining $\sim$60\% come from different target selection functions, mainly targeting luminous red galaxies \citep[LRGs; ][]{eisenstein01,dawson13}, have $\langle z_{\rm spec} \rangle = 0.48 \pm 0.13$, and represent by selection the most luminous galaxies at each redshift (e.g., see Figure~\ref{fig:template_color} and bottom panel of Figure~\ref{fig:zspecsample}).

Other spectroscopic programs targeting candidate globular clusters or UCDs (MMT, P.I. E. Peng; Peng et al., \textit{in preparation}; AAT, P.I.: P. C\^{o}t\'{e}: Zhang et al., \textit{submitted}; Zhang et al., \textit{in preparation}) provide us with $\sim$2,500 spec-z's ($\langle z_{\rm spec} \rangle = 0.16 \pm 0.12$).
We also gathered $\sim$90 spec-z's from the Virgo Dwarf Globular Cluster Survey taken with the Keck telescope (Keck, P.I.: P. Guhathakurta: Guhathakurta et al., \textit{in preparation}): those are faint ($i = 23.2 \pm 1.2$ mag) emission lines galaxies that were either purposely targeted (e.g., missclassified as GC, dwarf elliptical) or are serendipitous sources that happened to land on the "blank sky" portions of the slits.
Because of its faintness, this subsample is very different from the rest of our NGVSLenS spectroscopic subsamples and provides a unique opportunity to probe -- even sparsely -- our photo-z's up to $z_{\rm spec} \sim 1.5.$ for $21 < i < 24.5$ mag.

We display in Figure~\ref{fig:zspecsample} the spatial distribution (top panel) and $i$-band magnitude vs. $z_{\rm spec}$ distribution (bottom panel) of our spec-z compilation.
Our spec-z compilation over the NGVSLenS is very heterogeneous and rather shallow in redshift ($\sim$50\% with $z_{\rm spec} \lesssim 0.3$).
At low redshift ($z_{\rm spec} \lesssim 0.3$), it covers a wide range of colors and galaxy types; however at higher redshifts  ($z_{\rm spec} \gtrsim 0.3$), it is severely biased towards LRGs.

\subsubsection{Spectroscopic sample over the CFHTLenS}
To complement this spectroscopic sample, we use the public CFHTLenS data \citep{erben13} covering three intensive and deep spectroscopic surveys: the DEEP2 Galaxy Redshift Survey over the Extended Groth Strip \citep[DEEP2/EGS;][]{davis03,newman13a}, the VIMOS Public Extragalactic Redshift Survey \citep[VIPERS;][]{guzzo13}, and the F02 and F22 fields of the VIMOS VLT Deep Survey \citep[VVDS;][]{le-fevre05,le-fevre13}.
The DEEP2/EGS survey is a magnitude limited survey ($R \le 24.1$ mag, $\sim12,000$ galaxies), as is the VVDS ($17.5 \le i \le 24$ mag for the F02 field, $\sim$5,000 galaxies; $i \le 22.5$ mag for the F22 field, $\sim$4,000 galaxies), whereas the VIPERS ($\sim$30,000 galaxies) is color pre-selected and mainly targets objects in the range $0.5 \lesssim z_{\rm spec} \lesssim 1.2$ down to $i \sim 22.5$ mag.
For these three surveys, we select only galaxies having a secure redshift (flag $=\{3,4\}$ for the DEEP2/EGS and VVDS; $3 \le \textnormal{flag} < 5$ for the VIPERS).
Furthermore, the CFHTLenS fields covering these three deep spectroscopic surveys are also covered by the SDSS, which provides additional spec-z's ($\sim$6,400 galaxies, mostly obtained from surveys other that the Galaxy Main Sample).
In Figure~\ref{fig:zspecsample}, we present our CFHTLenS spectroscopic sample as black points.
It includes $\sim$57,000 spec-z's and spreads over $\sim$42 deg$^2$, which mitigates the field-to-field variations in exposure times.\\

\subsubsection{Photometric sample over the NGVSLenS}
To see how the properties of our spectroscopic samples compare to the NGVSLenS photometric data -- and to which extent they are representative thereof -- we define two photometric samples as follows.
We select objects: (1) lying in the 34 NGVSLenS fields having $u^*griz$-bands coverage (we exclude the overlap regions), (2) with valid photometry in those five bands, (3) not classified as star or GC using the criteria described in Appendix~\ref{sec:StarGCrem}.

We define NGVSLenS/phot23 (NGVSLenS/phot24, respectively) as the corresponding NGVSLenS photometric sample, when further applying a $i < 23$ mag ($i < 24$ mag, respectively) cut, which comprises $5.8 \times 10^5$ ($12.6 \times 10^5$, respectively) objects.\\

\begin{figure}[!h]
	\includegraphics[width=\columnwidth]{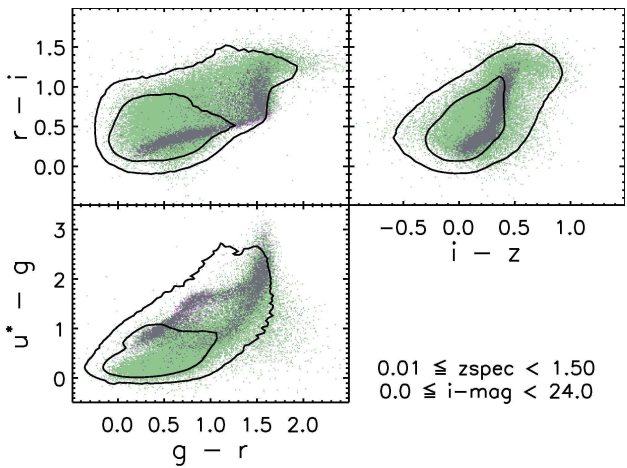}
	\caption{Coverage of the color-color space by our spectroscopic samples.
Magenta and green dots show our NGVSLenS and CFHTLenS spectroscopic samples, respectively.
Contours represent the 68\% and 95\% loci of our NGVSLenS/phot24 photometric sample.
Our spectroscopic samples satisfactorily cover the colors of the photometric sample.}
	\label{fig:zspec_colcov}
\end{figure}

In Figure~\ref{fig:zspec_colcov}, we show how our combined spectroscopic sample (over the NGVSLenS and the CFHTLenS) overlaps with the color-color space of the NGVSLenS data (NGVSLenS/phot24 sample).
The colored dots represent our spectroscopic sample and the contours are the 68\% and 95\% loci of the observed NGVSLenS photometric objects.
We remark that the "blue" sides (i.e.,  towards the bottom-left side) of the 95\% contours which are not well covered by our spectroscopic sample are mainly populated by $23 < i <24$ mag objects; in other words, our spectroscopic sample spans with high coverage the color-color space for $i < 23$ mag objects, and satisfactorily covers the $23 < i <24$ mag objects (the regions within the 68\% contours are well populated by our spectroscopic sample).

\subsection{Comparison with spec-z's}
\label{sec:zphot_zspec_ugriz}

We analyze in this section how our estimated photo-z's compare with our spectroscopic sample.
We use the full spectroscopic sample (NGVSLenS and CFHTLenS) in the redshift range $0.01 \le z_{\rm spec} < 1.5$, without any selection in magnitude.
For each object in our spectroscopic sample, we calculate $\Delta z = \frac{z_{\rm phot}-z_{\rm spec}}{1+z_{\rm spec}}$ and classify it as an outlier if $| \Delta z | > 0.15$.
For each considered sample, we report \textit{bias}: the median value of $\Delta z$; \textit{outl.}: the percentage of outliers; and $\sigma_{\rm outl.rej.}$: the standard deviation of $\Delta z$ when outliers have been excluded.
These quantities are used to facilitate comparison with other works; as mentioned in \citet{hildebrandt12}, the outlier definition is arbitrary.

\begin{figure}[!h]
	\includegraphics[width=\columnwidth]{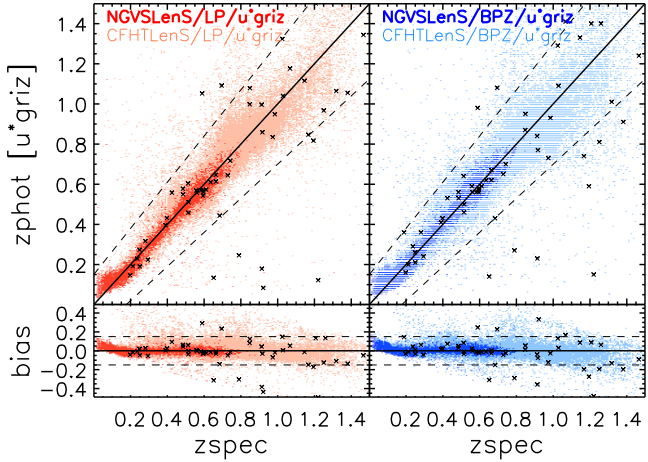}
	\caption{Photo-z's with $u^*griz$ bands with \textit{Le Phare} (\textit{left panel}, red) and \textsc{bpz} (\textit{right panel}, blue).
Dark thick symbols represent the NGVSLenS spectroscopic sample (low redshift); light thin symbols represent the CFHTLenS spectroscopic sample (high redshift).
We highlight with black crosses the NGVSLenS Keck subsample.}
\label{fig:zphotVSzspec_ugriz}
\end{figure}

We present in Figure~\ref{fig:zphotVSzspec_ugriz} how our photo-z's compare with spec-z's for our two spectroscopic samples and for both codes, \textit{Le Phare} (left panel, in red) and \textsc{bpz} (right panel, in blue).
The NGVSLenS objects (low redshift) are in thick dark symbols and the CFHTLenS objects (high redshift) are in thin light symbols.

At first sight, we see that both codes provide satisfactory photo-z's over the range $0.1 \lesssim z_{\rm spec} \lesssim 1$ and that the overall behavior of our spectroscopic samples over the NGVSLenS and the CFHTLenS fields is consistent in the overlap regions.
We notice that, although statistically small, the NGVSLenS Keck subsample has photo-z's in broad agreement with the other NGVSLenS subsamples and with the CFHTLenS spectroscopic sample, strengthening our choice of using the CFHTLenS data at high redshift.

\begin{figure}[!h]
	\includegraphics[width=\columnwidth]{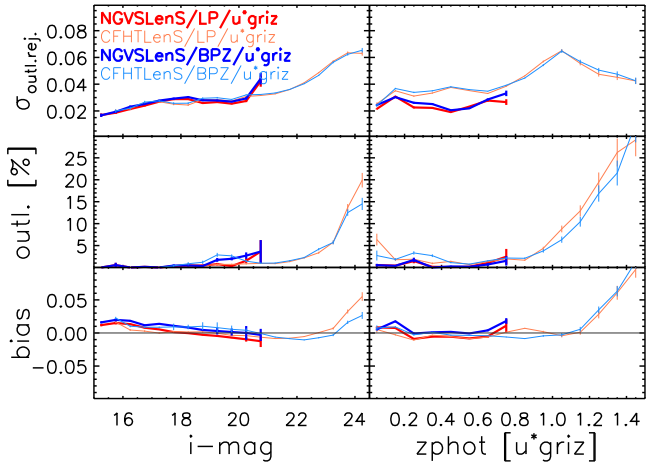}
	\caption{Statistics for photo-z's (estimated with $u^*griz$ bands) as a function of magnitude (\textit{left panel}) and redshift (\textit{right panel}).
Photo-z's estimated with \textit{Le Phare} are in red and those estimated with \textsc{bpz} are in blue.
Dark thick lines represent the NGVSLenS spectroscopic sample (low redshift); light thin lines represent the CFHTLenS spectroscopic sample (high redshift).
We report quantities only for the bins where we have more than 50 galaxies.
Error bars are calculated assuming a Poissonian distribution.
Our photo-z's estimated with the $u^*griz$-bands are more accurate for $i \lesssim 23$ mag or $z_{\rm phot} \lesssim 1$.}
\label{fig:stats_ugriz}
\end{figure}

We show in Figure~\ref{fig:stats_ugriz} a quantitative analysis of how the three quantities \textit{bias}, $\sigma_{\rm outl.rej.}$, and \textit{outl.} depend on the measured magnitude and the estimated photo-z.

\begin{figure}[!h]
	\includegraphics[width=\columnwidth]{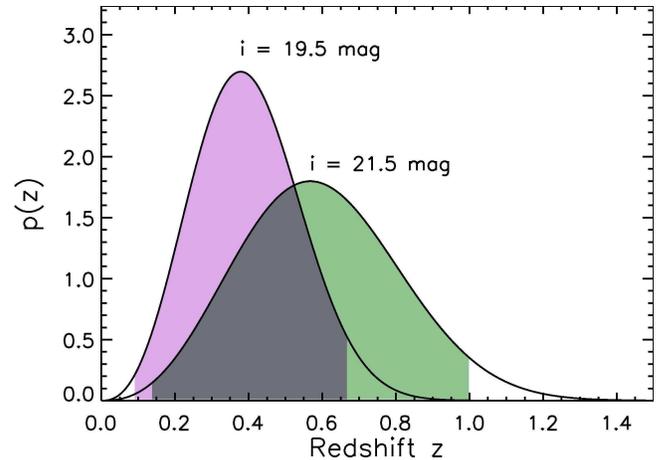}
	\caption{Example of the prior for an elliptical template for two magnitudes representative of our spectroscopic samples at a redshift of $z \sim 0.5$: $i = 19.5$ mag (in magenta, representative of our NGVSLenS spectroscopic sample) and $i = 21.5$ mag (in green, representative of our CFHTLenS spectroscopic sample).
The shaded areas enclose 95\% of the prior.
As a consequence, brighter galaxies will have on average sharper posteriors.}
\label{fig:prior}
\end{figure}

First, we observe that the two codes (\textsc{bpz} and \textit{Le Phare}) and the two datasets (NGVSLenS and CFHTLenS) provide consistent behavior over our tested ranges in magnitude or photo-z. 
This observation \textit{a posteriori} validates our assumption, namely that the NGVSLenS and CFHTLenS data have very similar properties.
However, when looking at the $\sigma_{\rm outl.rej.}$ as a function of photo-z, there is a clear difference between the NGVSLenS and CFHTLenS samples, which arises from the nature of the two spectroscopic samples in a given range.
In the $0.3 \lesssim z_{\rm phot} \lesssim 0.6$ range, the NGVSLenS sample has a significantly smaller $\sigma_{\rm outl.rej.}$.
This is a direct consequence being of our NGVSLenS spectroscopic sample in this redshift range is highly biased towards LRGs.
These galaxies have on average brighter magnitudes and smaller photometric errors (e.g., better defined  4,000 \AA~ break than the average galaxy), thus making the photo-z estimation easier. 
For instance, the typical value of the $i$-band magnitude at redshift $\sim$0.5 is 19.5 mag for our NGVSLenS spectroscopic sample versus 21.5 mag for our CFHTLenS spectroscopic sample (see right panel of Figure~\ref{fig:zspecsample}): a direct consequence is that the prior for those LRGs is significantly more peaked and at lower redshifts, thus constraining more the posterior.
Indeed, as illustrated in Figure~\ref{fig:prior}, the [0.11,0.65] redshift interval includes 95\% of the prior for an elliptical galaxy with $i = 19.5$ mag, whereas for an elliptical galaxy with $i = 21.5$ mag the corresponding redshift interval is [0.14,1.00].

Another trend that illustrates this point is the outlier rate at $z_{\rm phot} \lesssim 0.2$: the CFHTLenS sample has 3-5\% outliers against $<$1\% for the NGVSLenS sample.
Again, this can be explained by the characteristic $i$-band magnitude in this redshift range: when considering objects with $z_{\rm phot} < 0.2$, only 5\% of the NGVSLenS sample has $i > 20$ mag against 27\% for the CFHTLenS sample.
The fainter galaxies of the CFHTLenS will have a much broader prior, hence the photo-z will be less constrained.

The quality of our photo-z's decreases with increasing magnitude or redshift.
For instance, with both codes, the $bias$ becomes significant ($>0.02$) for faint ($i \gtrsim 23$ mag) or high-z ($z \gtrsim 1.2$) objects,
and the $\sigma_{\rm outl.rej.}$ goes from $\sim$0.02 for bright/low-z objects to $\sim$0.06 for faint/high-z objects.
For $z_{\rm phot} \gtrsim 1.2$, our optical data do not bracket  the 4,000 \AA~ break, and the photo-z's are less reliable.

The quality of our photo-z  for $i \gtrsim 17.5$ mag is consistent with that from the CFHTLenS data computed with \textit{Le Phare} \citep{ilbert06a, coupon09} or with \textsc{bpz} \citep{hildebrandt12,erben13}.
However, we obtain more robust photo-z's down to at least $i \simeq 15.5$ mag because of the new prior that we introduce for the brightest galaxies.

\subsection{Individual photo-z uncertainty $z_{\rm phot,err.}$}
\label{sec:zphoterr}

In this section we discuss the individual photo-z uncertainty estimation, which we label $z_{\rm phot,err.}$.
For \textit{Le Phare}, we use \texttt{Z\_ML68\_LOW} and \texttt{Z\_ML68\_HIGH}, which represent for each galaxy the boundary of the interval including 68\% of the redshift probability distribution function;
\textsc{bpz} provides only 95\% uncertainty on $z_{\rm phot}$: for each object, we compute, based on the output posterior, the 68\% confidence interval as defined in \textit{Le Phare}.

\begin{figure}[!h]
	\includegraphics[width=\columnwidth]{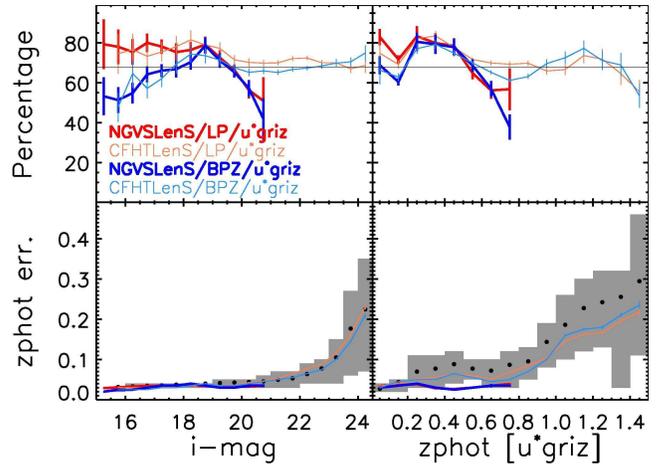}
\caption{Properties of, $z_{\rm phot,err.}$, the individual error estimation on the photo-z (estimated with $u^*griz$ bands) as a function of magnitude (\textit{left panel}) and redshift (\textit{right panel}).
Photo-z's estimated with \textit{Le Phare} are in red and those estimated with \textsc{bpz} are in blue.
Dark thick lines represent the NGVSLenS spectroscopic sample (low redshift); light thin lines represent the CFHTLenS spectroscopic sample (high redshift).
Error bars are calculated assuming a Poissonian distribution.
\textit{Top panels}: percentage of objects having $z_{\rm spec}$ within $z_{\rm phot} \pm z_{\rm phot,err.}$. The fact that these percentages are close to $68\%$, as expected from the $z_{\rm phot,err.}$ uncertainty estimation, means that our estimated $z_{\rm phot,err.}$ are realistic.
\textit{Bottom panels}: median value of $z_{\rm phot,err.}$ (the continuous red and blue line are for  \textit{Le Phare}  and \textsc{bpz}, respectively); for each bin along the x-axis, the black dots represent the median value of $z_{\rm phot,err.}$ for photometric objects in our NGVSLenS/phot24 sample and the grey shaded areas represent the regions enclosing $68\%$ of their distribution.
Our spectroscopic sample is representative of the photometric sample.}
\label{fig:zphoterr_ugriz}
\end{figure}

In the top panels of Figure~\ref{fig:zphoterr_ugriz}, we present the percentage of objects having $z_{\rm spec}$ within $z_{\rm phot} \pm z_{\rm phot,err.}$, for our spectroscopic sample and as a function of measured magnitude or photo-z.
On average, this percentage is close to $68\%$, which means that our estimated individual $z_{\rm phot,err.}$ are realistic.

We see a departure from this behavior for the NGVSLenS sample at $i \gtrsim 20$ mag and at $z_{\rm phot} \gtrsim 0.5$: this is again due to the fact that our NGVSLenS spectroscopic sample in this magnitude/redshift range is highly biased towards LRGs.
Indeed, as those objects are bright with a clear 4,000 \AA~ break, the estimated individual $z_{\rm phot,err.}$ is small, hence the $z_{\rm spec}$ can be outside of the $z_{\rm phot} \pm z_{\rm phot,err.}$ interval.
This means that the uncertainties obtained for the photometric redshift are underestimated for LRGs: for this population, setting a minimal value of 0.04 for $z_{\rm phot,err.}$ allows to recover realistic errors.\\

In the lower panels of Figure~\ref{fig:zphoterr_ugriz} we present how the median value of $z_{\rm phot,err.}$ varies as a function of the measured magnitude and $z_{\rm phot}$.
The median value of $z_{\rm phot,err.}$ depends strongly on magnitude.
The grey area shows the region including the  68\% of our NGVSLenS/phot24 sample photometric sample.
For the spectroscopic sample,  the median value of $z_{\rm phot,err.}$ is comparable to the scatter $\sigma_{\rm outl.,rej.} \times (1+z_{\rm spec})$ for $i < 23$ mag or $z_{\rm phot} < 1$ (i.e., when the factor $(1+z_{\rm spec})$ is taken into account).
Our spectroscopic sample is representative of the overall behavior of all photometric objects, even if  the spectroscopic sample has lower median uncertainties as a function of redshift, because the majority of the photometric sample includes objects fainter than the spectroscopic sample ($i > 22$ mag).

\subsection{Photo-z's without the $r$-band}
\label{sec:zphot_nor}

As mentioned in Section~\ref{sec:data}, a majority (83/117 fields) of the NGVSLenS field has not been imaged yet with the $r$-band.
In this Section, we present the quality of our photo-z's when the $r$-band is not available.
To estimate it, we re-calculated the photo-z's using only the $u^*giz$-bandpasses for the 34 NGVSLenS fields having $r$-band coverage, and for the CFHTLenS fields, thus using our full spectroscopic sample ($\sim$83,000 galaxies).

\begin{figure}[!h]
	\includegraphics[width=\columnwidth]{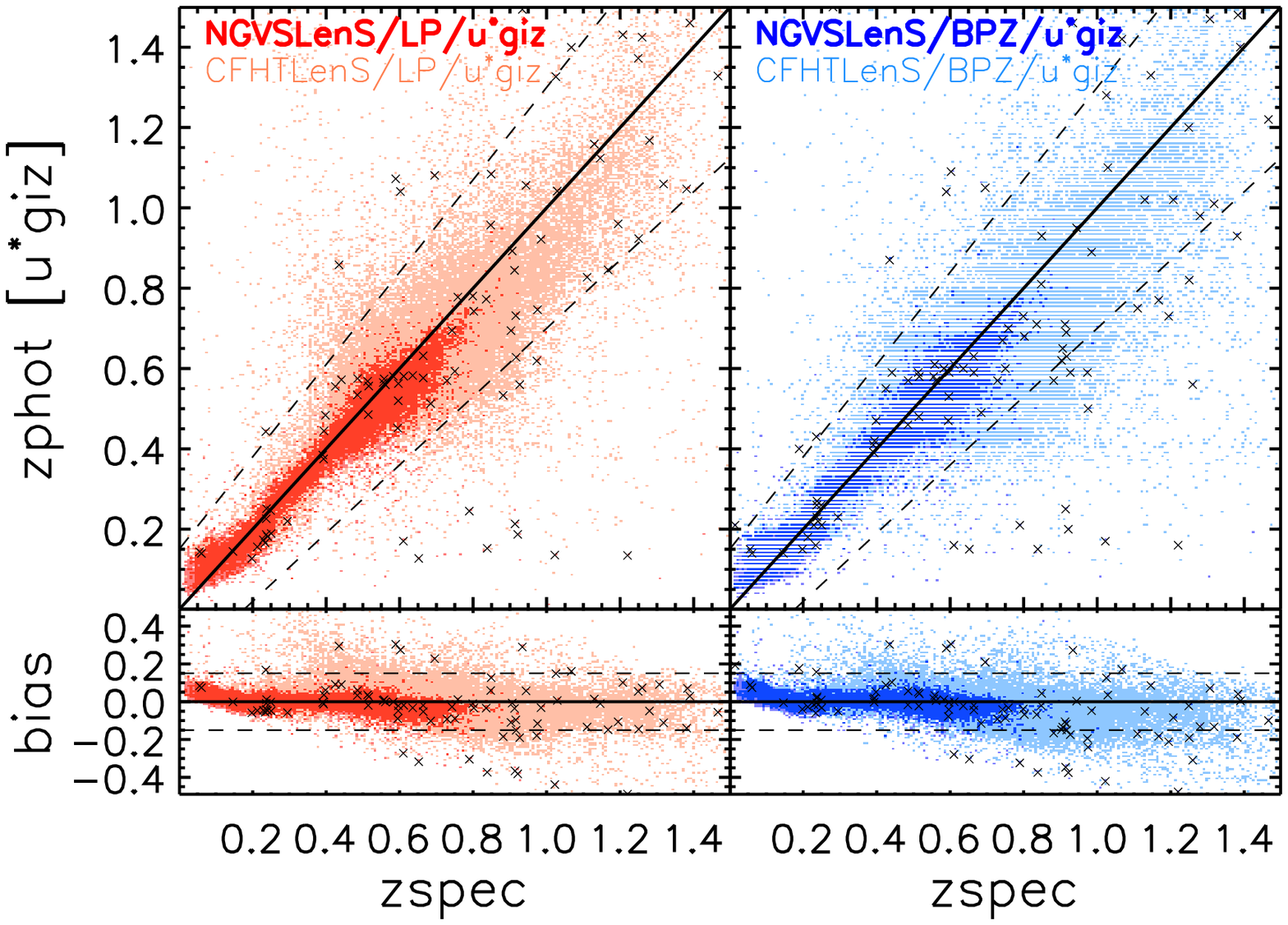}
	\caption{Photo-z's with $u^*giz$ bands with \textit{Le Phare} (\textit{left panel}, red) and \textsc{bpz} (\textit{right panel}, blue).
Dark thick symbols represent the NGVSLenS spectroscopic sample (low redshift); light thin symbols represent the CFHTLenS spectroscopic sample (high redshift).
We highlight with black crosses the NGVSLenS Keck subsample.}
\label{fig:zphotVSzspec_ugiz}
\end{figure}

\begin{figure}[!h]
	\includegraphics[width=\columnwidth]{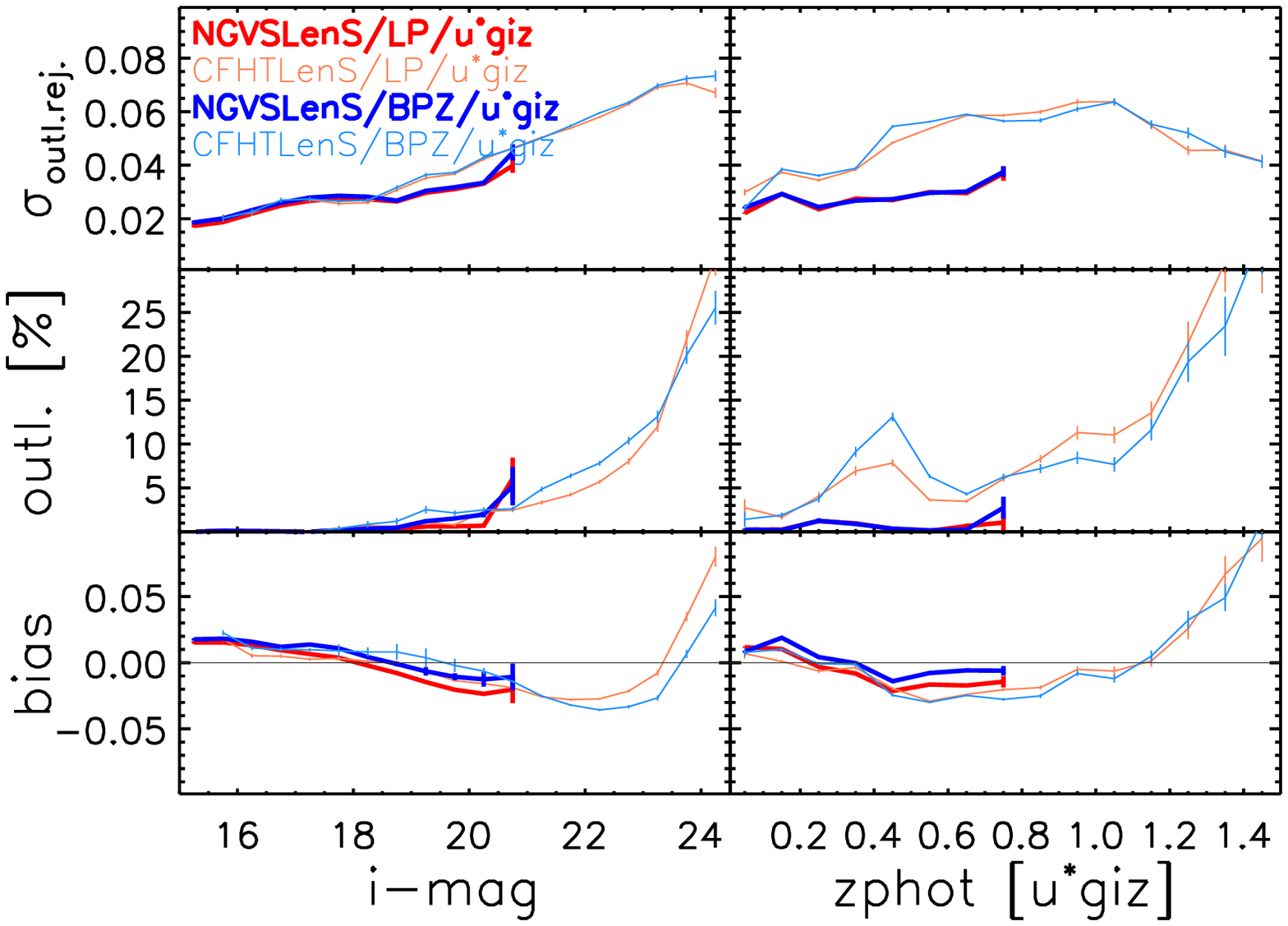}
	\caption{Statistics for photo-z's (estimated with $u^*giz$ bands) as a function of magnitude (\textit{left panel}) and redshift (\textit{right panel}).
Photo-z's estimated with \textit{Le Phare} are in red and those estimated with \textsc{bpz} are in blue.
Dark thick lines represent the NGVSLenS spectroscopic sample (low redshift); light thin lines represent the CFHTLenS spectroscopic sample (high redshift).
We report quantities only for the bins where we have more than 50 galaxies.
Error bars are calculated assuming a Poissonian distribution.
When the $r$-band is missing the quality of our photo-z's decreases for the three plotted quantities in the $0.3 \lesssim z_{\rm phot} \lesssim 0.8$ range and in the $i \gtrsim 21$ mag range (\textit{bias}, \textit{outl.}, and $\sigma_{\rm outl.rej.}$ can increase of more than 100\%).}
\label{fig:stats_ugiz}
\end{figure}

\begin{figure*}[!h]
	\begin{tabular}{c @{\hskip 30pt} c}
		\includegraphics[width=\columnwidth]{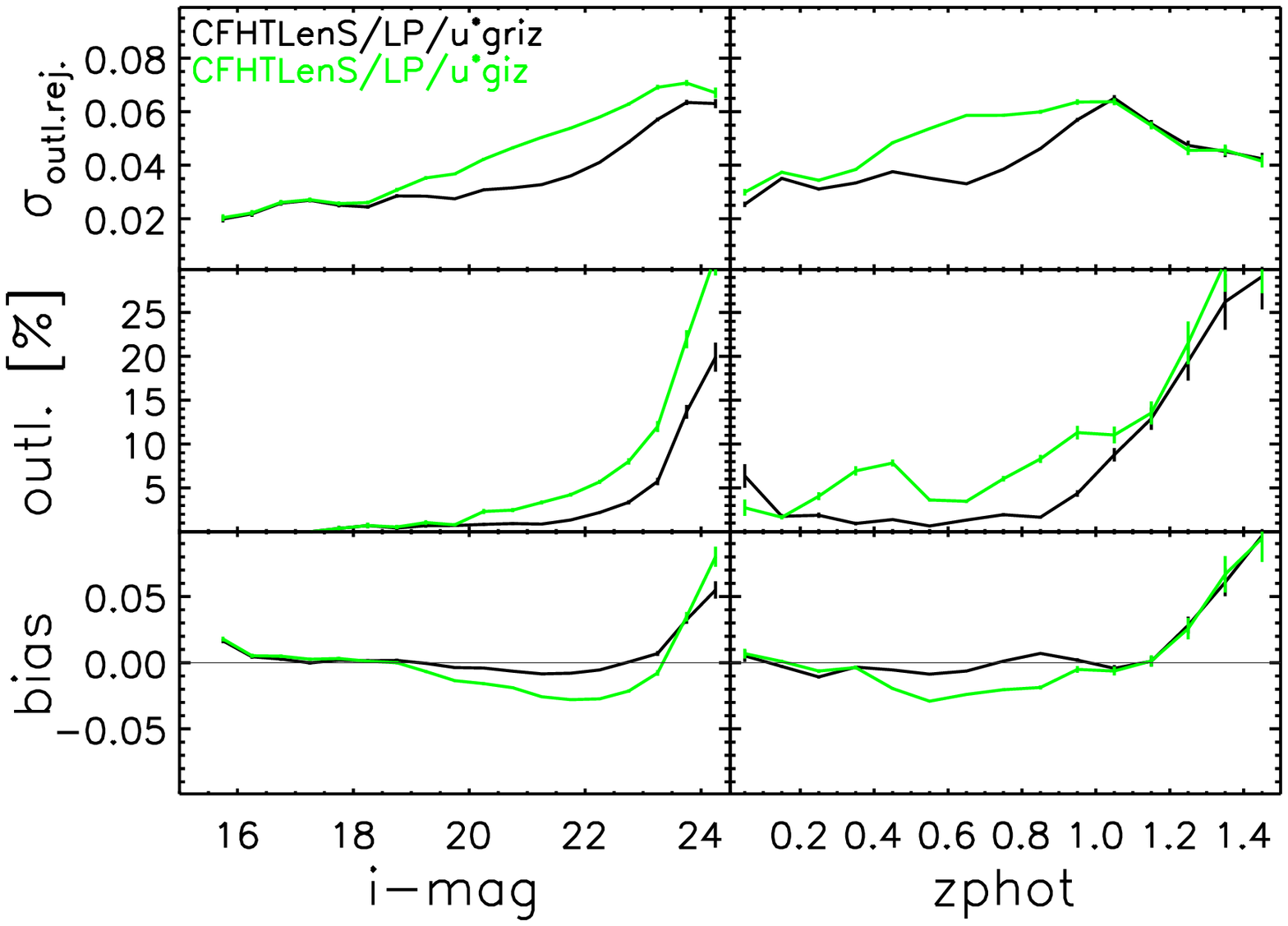} &
		\includegraphics[width=\columnwidth]{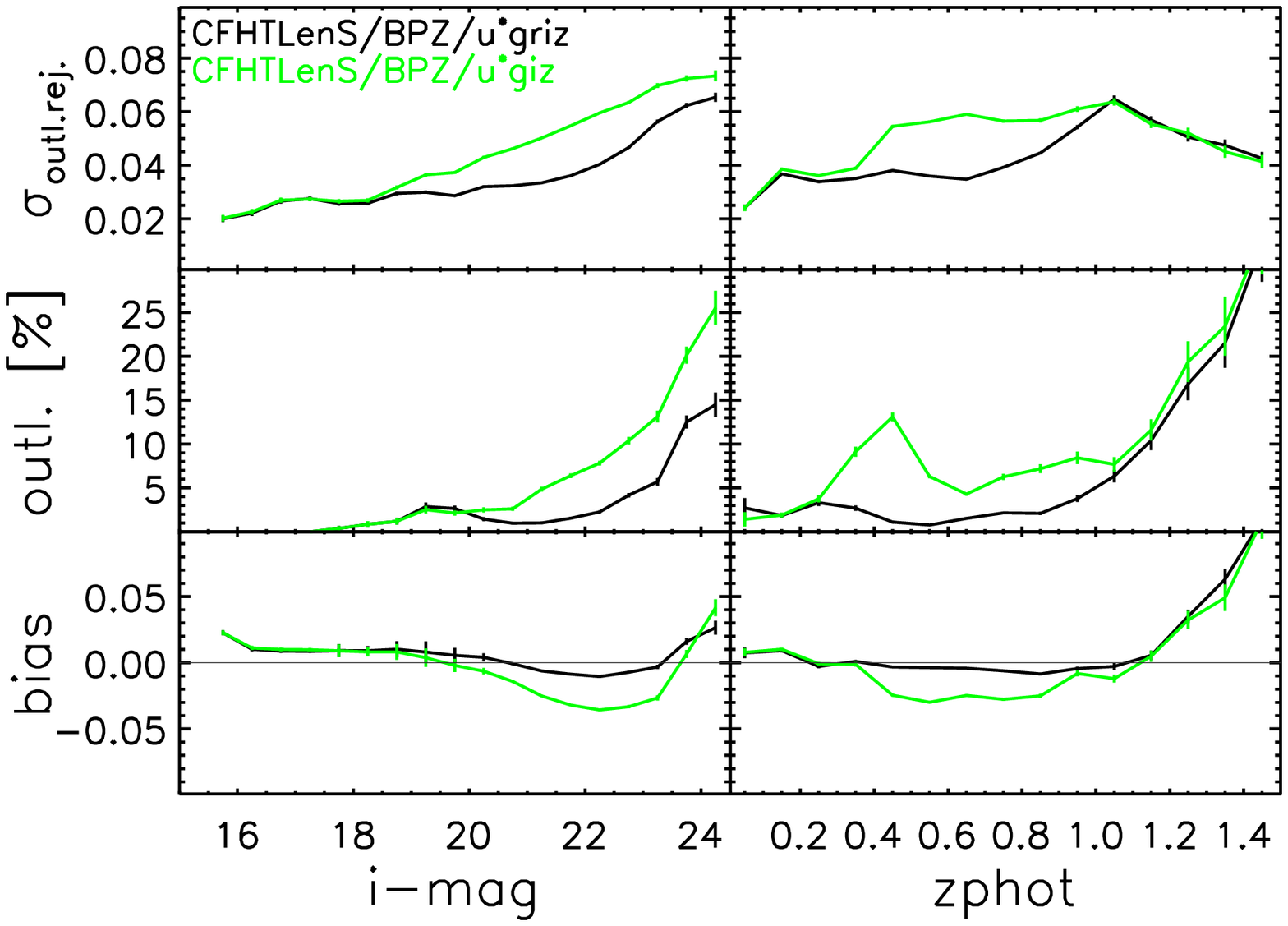} \\
	\end{tabular}
	\caption{Statistics for photo-z's estimated with \textit{Le Phare} (left) and with \textsc{bpz} (right), as a function of magnitude and redshift, for our CFHTLenS spectroscopic sample.
Photo-z's estimated with the $u^*griz$ bands are in black and those estimated with the $u^*giz$ bands are in green.
We report quantities only for the bins where we have more than 50 galaxies.
Error bars are calculated assuming a Poissonian distribution.
When the $r$-band is missing the quality of our photo-z's decreases for the three plotted quantities in the $0.3 \lesssim z_{\rm phot} \lesssim 0.8$ range and in the $i \gtrsim 21$ mag range (\textit{bias}, \textit{outl.}, and $\sigma_{\rm outl.rej.}$ can increase of more than 100\%).}
\label{fig:stats_ugriz_ugiz}
\end{figure*}

Figs. \ref{fig:zphotVSzspec_ugiz} and \ref{fig:stats_ugiz} summarize the properties of the photo-z's for our spectroscopic sample when the $r$-band is missing.
We also present in Figure \ref{fig:stats_ugriz_ugiz} the comparison of the statistics for the photo-z's estimated with or without the $r$-band for our CFHTLenS spectroscopic sample.
In our CFHTLenS spectroscopic sample, our photo-z's are more scattered in the $0.3 \lesssim z_{\rm spec} \lesssim 0.8$ range, where the $r$-band filter is essential to constrain the 4,000 \AA~ break.
This effect is less pronounced for our NGVSLenS spectroscopic sample because, as discussed in Section~\ref{sec:zphot_zspec_ugriz} (see also Figure~\ref{fig:prior}), our NGVSLenS spectroscopic sample is highly biased towards LRGs (i.e., thus not representative of the general galaxy population) in this redshift range: the prior -- more peaked and at lower redshift than for average galaxies at similar redshift -- helps to obtain fewer false values for the posterior.
When we compare the statistics, (e.g., Figure~\ref{fig:stats_ugiz} with Figure~\ref{fig:stats_ugriz}), when the $r$-band is missing the outliers rate increases significantly in the $0.3 \lesssim z_{\rm phot} \lesssim 0.8$ range and peaks at $\sim$10-15\%. 
We remark that the overall behavior of the two codes is similar.\\

The observation that the photo-z quality decreases in the $0.3 \lesssim z_{\rm spec} \lesssim 0.8$ range is supported only by our CFHTLenS spectroscopic sample.
Using the results of Section~\ref{sec:zphot_zspec_ugriz}, we can assume that our photo-z's estimated with five bands $u^*griz$ are unbiased down to $i < 23$ mag.
Under this assumption, we can use our NGVSLenS/phot23 sample (see Section~\ref{sec:samples} and Table \ref{tab:samples}), i.e.,  the whole NGVSLenS photometric sample with $i < 23$ mag and covered by the five bands, to probe how photometric redshifts change if the $r$-band is missing.

\begin{figure}[!h]
	\includegraphics[width=\columnwidth]{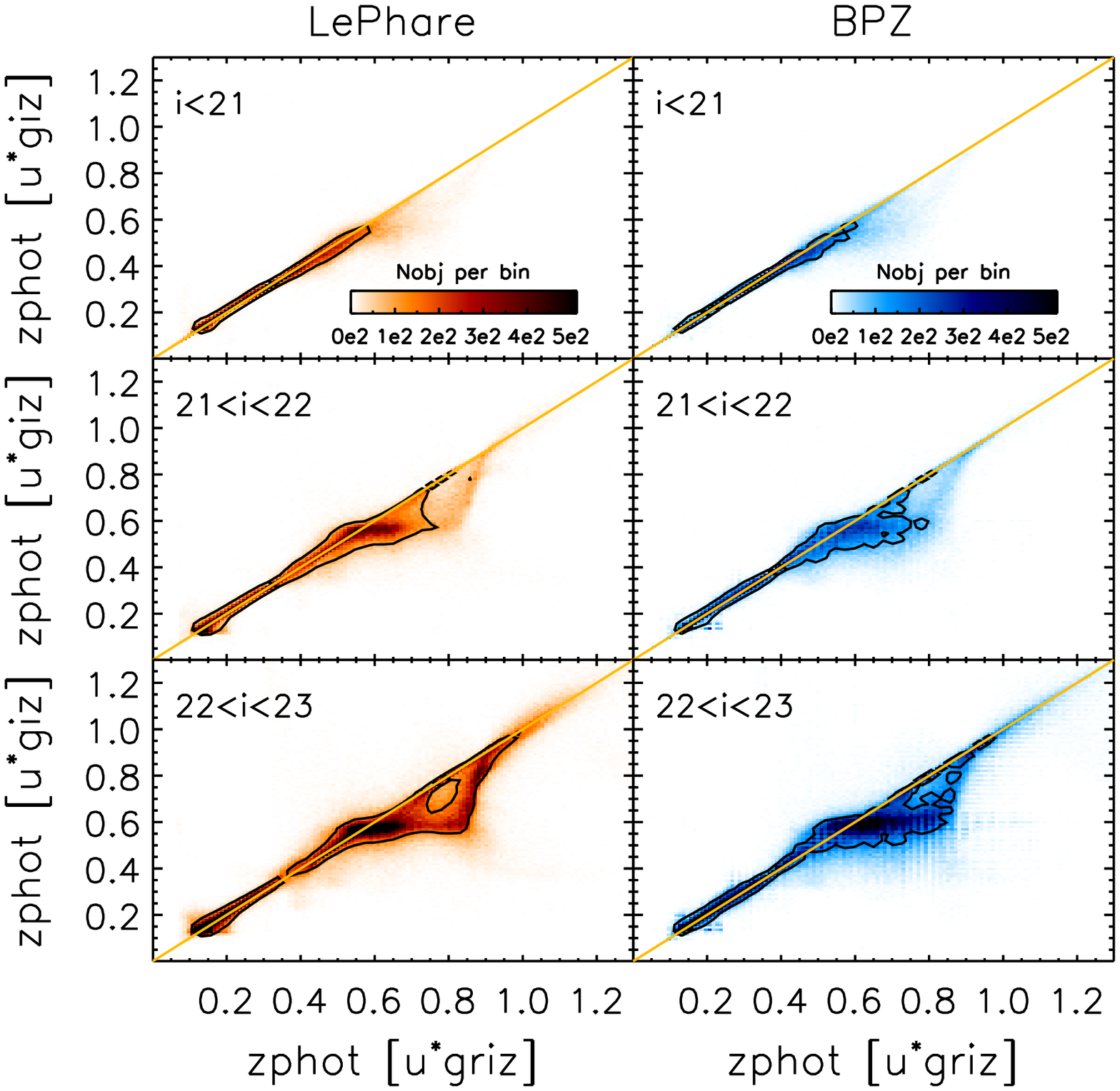}
\caption{Photo-z's with $u^*giz$ bands vs. photo-z's with $u^*griz$ bands for our NGVSLenS/phot23 photometric sample, by magnitude bins.
\textit{Left column}: photo-z's estimated with \textit{Le Phare}; \textit{right column}: photo-z's estimated with \textsc{bpz}.
We represent the density of objects per 0.01$\times$0.01 bin in $z_{\rm phot}$.
Black contours represent the loci enclosing 68\% of the distribution.
The analysis with a large NGVSLenS photometric sample confirms the one done with the CFHTLenS spectroscopic sample.}
\label{fig:ugiz_vs_ugriz}
\end{figure}

Figure~\ref{fig:ugiz_vs_ugriz} compares the photo-z's estimated with $u^*giz$ bands versus those estimated with the $u^*griz$ bands, for the two codes and by magnitude bins.
We consider here objects from our NGVSLenS/phot23 photometric sample, thus $5.8 \times 10^5$ objects.
It confirms the result obtained with the CFHTLenS spectroscopic sample, which is that a significant number of objects with $0.3 \lesssim z_{\rm phot} \lesssim 0.8$ are outliers. 

\begin{figure}[!h]
	\includegraphics[width=\columnwidth]{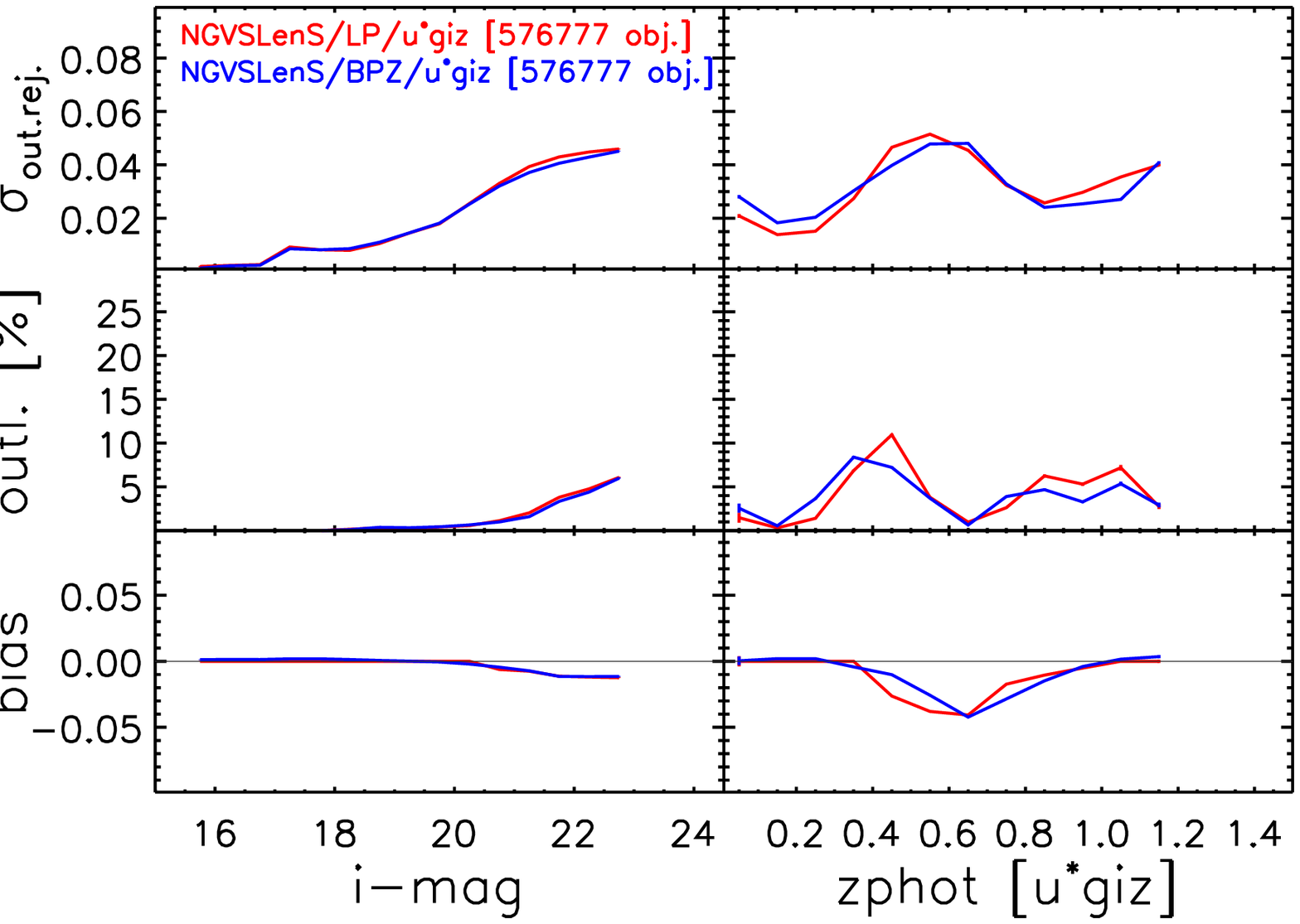}
	\caption{
Statistics for photo-z's (estimated with $u^*giz$ bands) in our NGVSLenS/phot23 photometric sample ($>$$5 \times 10^5$ objects), as a function of magnitude (\textit{left panel}) and redshift (\textit{right panel}).
Photo-z's estimated with \textit{Le Phare} are in red and those estimated with \textsc{bpz} are in blue.
Figure similar to Figure~\ref{fig:stats_ugiz}, but the spectrospic sample has been replaced by a NGVSLenS photometric sample ($5.8 \times 10^5$ objects) and the spec-z's have been replaced by the photo-z's estimated with the $u^*griz$ bands.
Error bars, calculated assuming a Poissonian distribution, are not visible because of the large size of the sample.
The analysis with a large NGVSLenS photometric sample confirms the one done with the CFHTLenS spectroscopic sample.}
\label{fig:nor_photstats}
\end{figure}

Under the same assumption -- our photo-z's estimated with the $u^*griz$ bands are globally unbiased down to $i < 23$ mag -- we can produce a plot similar to Figure~\ref{fig:stats_ugiz}, but using our NGVSLenS/phot23 photometric sample, as above.
In Figure~\ref{fig:nor_photstats}, we present the statistics for our photo-z's estimated with the $u^*giz$ bands, using the photo-z's estimated with the $u^*griz$ bands as a proxy of $z_{\rm spec}$.
Most of the features of Figure~\ref{fig:stats_ugiz} are rather accurately reproduced with this sample of $5.8 \times 10^5$ photometric objects.
These statistics are dominated by faint galaxies, and less biased by the LRGs, and are consistent with statistics obtained with the CFHTLenS spectroscopic sample.

Figure~\ref{fig:nor_photstats} strengthens the results shown in Figure~\ref{fig:stats_ugiz}, as the statistics are closely reproduced by using a sample of $5.8 \times 10^5$ photometric objects.

\subsection{Joint analysis of photo-z dependence on magnitude and redshift}
\label{sec:joint}

The above analysis shows that the photo-z's statistics can be biased by the properties of the spectroscopic sample chosen as reference.
Indeed, the galaxies selected as targets for spectroscopic samples might have properties that are not characteristic of the entire photometric sample.
In particular, spectroscopic samples are dominated by brighter galaxies that represent a small percentage of the entire photometric samples, at a given redshift.
This means that fainter galaxies, which statistically dominate the photometric sample, will not be correctly represented in the spectroscopic sample.
In our case, we do have spectroscopic  samples that cover the fainter magnitudes (see Figure~\ref{fig:zspecsample}, bottom panel), however at a given redshift they include significantly fewer galaxies than those that cover the bright end, which dominate our statistics shown in previous sections. 

Moreover, as shown in Figure~\ref{fig:zspecsample},  our spectroscopic samples cover different ranges in redshift and magnitude and, when binning only  in magnitude \textit{or} in redshift, we are mixing galaxies that have very different properties. 

For those reasons, we here separate different magnitude bins at a given redshift, and perform a joint analysis in enough small bins of redshift \textit{and} magnitude, in which we can select galaxies with similar properties. \\

\begin{figure*}
	\vspace{10pt}
	\begin{tabular}{c @{\hskip 30pt} c}
	\includegraphics[width=\columnwidth]{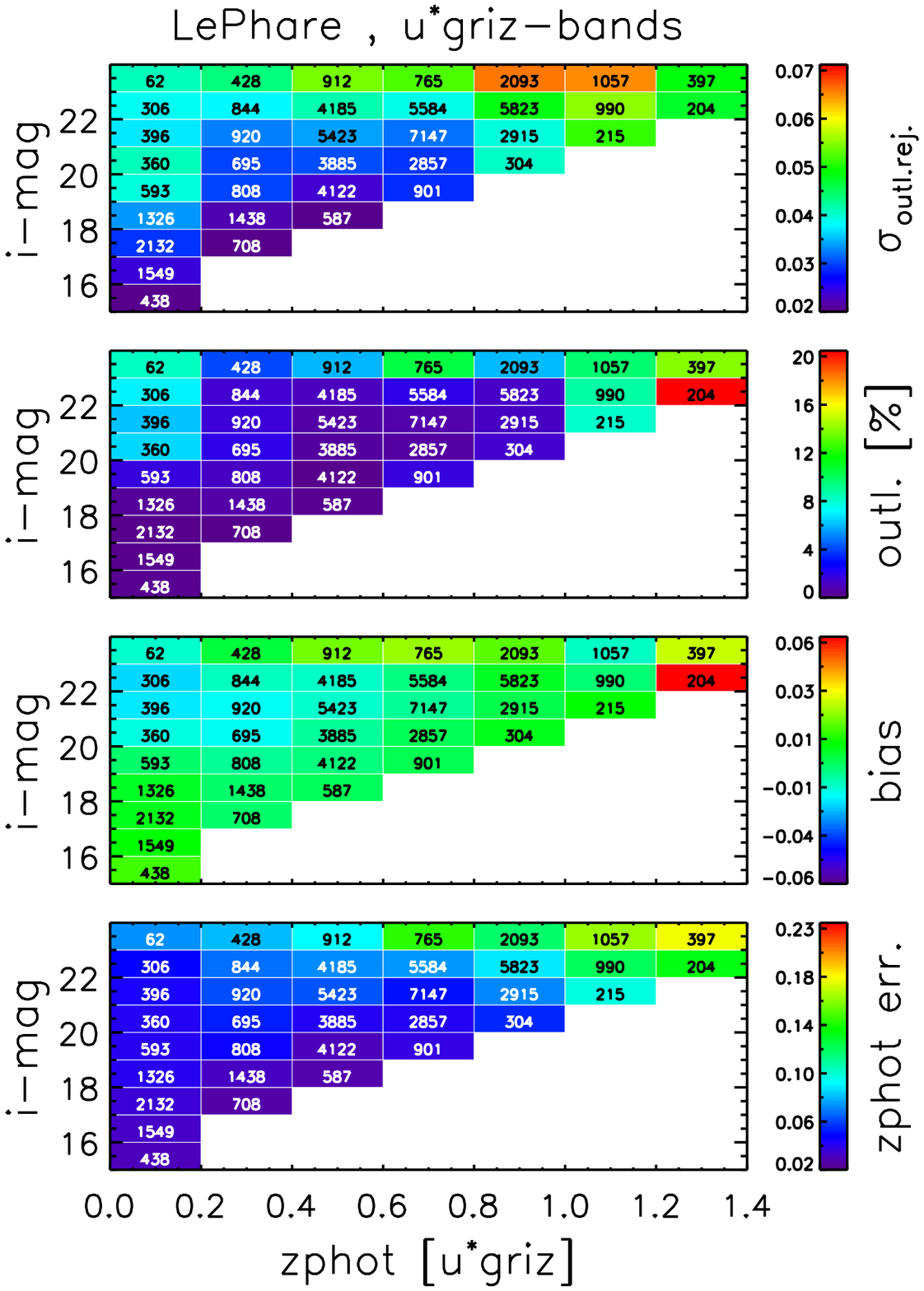} &
	\includegraphics[width=\columnwidth]{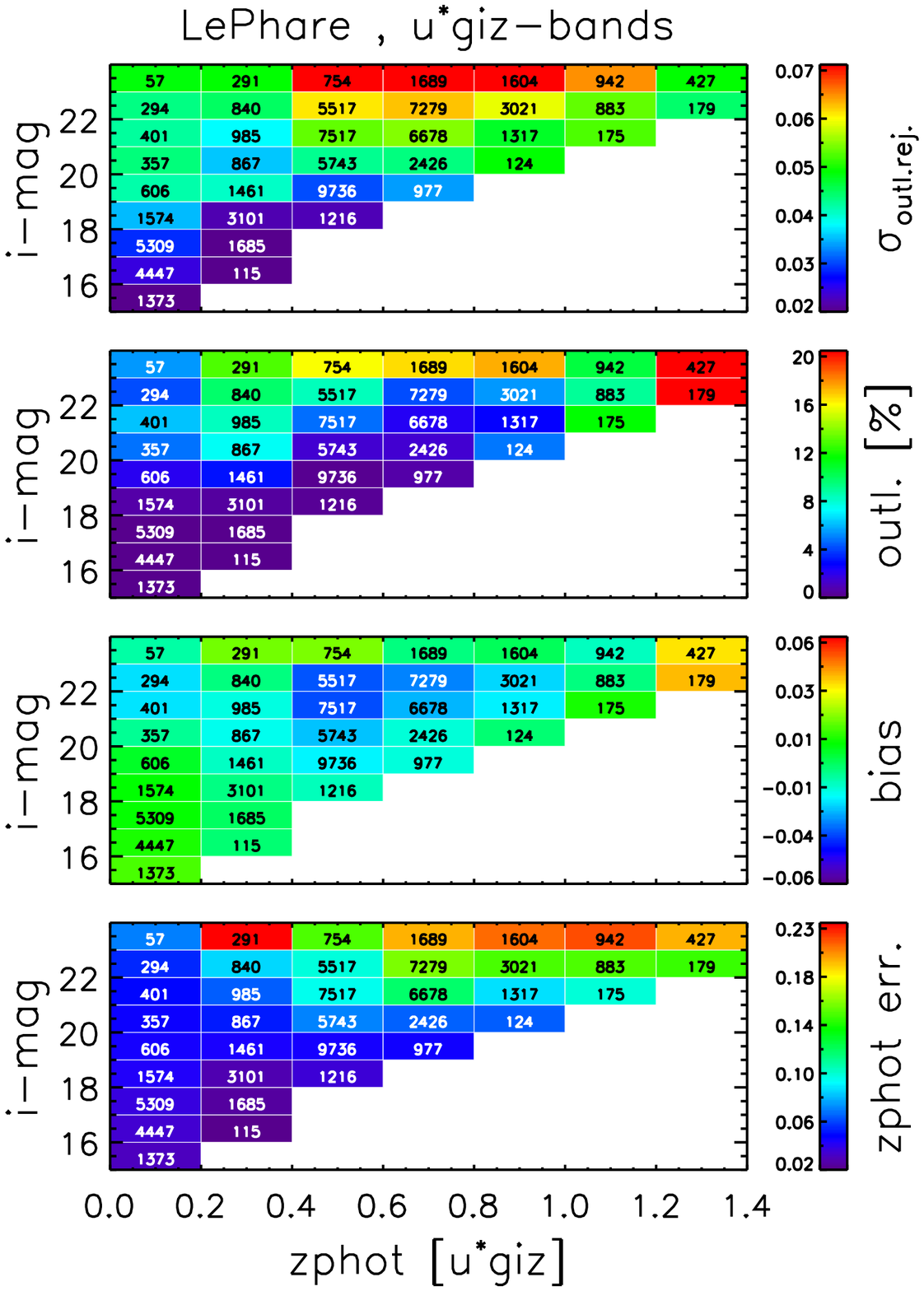} \\
	\end{tabular}
\caption{Joint analysis of our photo-z's estimated with \textit{Le Phare} and the $u^*griz$-bands (left panel) and the $u^*giz$-bands (right panel).
From top to bottom, we display $\sigma_{\rm outl.rej.}$, \textit{outl.}, \textit{bias}, and the median value of $z_{\rm phot,err.}$.
We use here our full spectroscopic sample (NGVSLenS \textit{and} CFHTLenS; $\sim$83,000 galaxies).
The statistics are presented by binning it \textit{both} in magnitude \textit{and} photo-z.
We only display the bins where we have more than 50 galaxies and, for those bins, we report the number of galaxies used to compute the statistics.\\
}
\label{fig:stats_crossbin}
\end{figure*}

In Figure~\ref{fig:stats_crossbin} and in Table \ref{tab:stats_crossbin}, we present a joint analysis in redshift and magnitude on the entire NGVSLenS and CFHTLenS spectroscopic samples ($\sim$83,000 galaxies). We have already discussed that \textit{Le Phare} and \textsc{bpz} provide photo-z's with similar properties and, for simplicity, we present only \textit{Le Phare} photo-z's for this analysis.
When using \textsc{bpz}, our results do not change.

This analysis clarifies what has been observed in the previous sections and refines the conclusions.
In a given redshift bin, the quality of the photo-z's depends on magnitude, with fainter galaxies having more uncertain photo-z's estimates.
Bright galaxies with $i < 21$ mag have accurate photo-z's, independent of redshift.
Their bias and scatter are low ($|bias|<0.02$ and $\sigma_{\rm outl.rej.}<0.04$) and they have a small number of outliers ($<6\%$).
This also true when the photo-z's are estimated without the $r$-band for $i < 20$ mag.
For fainter galaxies ($i>21$~mag), it results in a steeper decline in the accuracy of the photo-z estimate as a function of magnitude in the $0.3 \lesssim z_{\rm phot} \lesssim 0.8$.
As previously explained, this depends on tighter priors on the brightest galaxies, and on the predominance of galaxies with more defined 4,000 \AA~ breaks at the higher end of the luminosity function.

At fixed magnitude, photo-z estimates at higher redshift are often more accurate, just because we are probing galaxies with higher absolute luminosity, e.g. intrinsically brighter galaxies at higher redshift, that will have more defined 4,000 \AA~ breaks.
When the optical bandpasses no longer bracket the 4,000 \AA~ break ($z \gtrsim 1.2$), this is no longer true and the photo-z's become more uncertain even for bright red sequence galaxies.

\begin{deluxetable*}{llll | cccccc | cccccc}
	\tablewidth{0pt}
	\setlength{\tabcolsep}{3pt}
	\tabletypesize{\scriptsize}
	\tablecolumns{16}
	\tablecaption{Photo-z (estimated with \textit{Le Phare}) properties for our compiled spectroscopic sample (NGVSLenS and CFHTLenS). \label{tab:stats_crossbin}}
	\tablehead{
		\multicolumn{4}{c|}{Binning} &\multicolumn{6}{c|}{$u^*griz$-bands} & \multicolumn{6}{c}{$u^*giz$-bands}\\[2pt]
		\hline
		\colhead{$i^{\rm min}$} & \colhead{$i^{\rm max}$} &
			\colhead{$z_{\rm phot}^{\rm min}$} & \multicolumn{1}{c|}{$z_{\rm phot}^{\rm max}$} &
			\colhead{\textit{bias}} & \colhead{$\sigma_{\rm outl. rej.}$} & \colhead{\textit{outl.}}  &
			\colhead{$z_{\rm phot, err.}$} & \colhead{Ngal} & \multicolumn{1}{c|}{zSurvey$^\dag$} &
			\colhead{\textit{bias}} & \colhead{$\sigma_{\rm outl. rej.}$} & \colhead{\textit{outl.}} &
			\colhead{$z_{\rm phot, err.}$} & \colhead{Ngal} & \colhead{zSurvey$^\dag$}\\
		\colhead{[mag]} & \colhead{[mag]} & \colhead{} & \multicolumn{1}{c|}{} &
			\colhead{} & \colhead{} & \colhead{[\%]} & \colhead{}  & \colhead{} & \multicolumn{1}{c|}{} &
			\colhead{} & \colhead{} & \colhead{[\%]} & \colhead{}  & \colhead{} & \colhead{}
		}
	\startdata
15.0 & 16.0 & 0.0 & 0.2 & 0.01 & 0.02 & 0 & 0.03 & 438 & A & 0.02 & 0.02 & 0 & 0.03 & 1373 & A\\
16.0 & 17.0 & 0.0 & 0.2 & 0.01 & 0.02 & 0 & 0.03 & 1549 & A & 0.01 & 0.02 & 0 & 0.03 & 4447 & A\\
16.0 & 17.0 & 0.2 & 0.4 & \nodata & \nodata & \nodata & \nodata & \nodata & \nodata & 0.00 & 0.02 & 0 & 0.02 & 115 & A\\
17.0 & 18.0 & 0.0 & 0.2 & 0.01 & 0.03 & 0 & 0.04 & 2132 & A & 0.01 & 0.03 & 0 & 0.04 & 5309 & A\\
17.0 & 18.0 & 0.2 & 0.4 & 0.00 & 0.02 & 0 & 0.03 & 708 & D & 0.00 & 0.02 & 0 & 0.02 & 1685 & D\\
18.0 & 19.0 & 0.0 & 0.2 & 0.00 & 0.03 & 0 & 0.04 & 1326 & B & 0.01 & 0.03 & 1 & 0.04 & 1574 & B\\
18.0 & 19.0 & 0.2 & 0.4 & 0.00 & 0.02 & 1 & 0.03 & 1438 & D & -0.01 & 0.02 & 0 & 0.03 & 3101 & D\\
18.0 & 19.0 & 0.4 & 0.6 & 0.00 & 0.02 & 0 & 0.03 & 587 & D & -0.01 & 0.02 & 0 & 0.04 & 1216 & D\\
19.0 & 20.0 & 0.0 & 0.2 & 0.00 & 0.04 & 2 & 0.04 & 593 & C & 0.00 & 0.04 & 2 & 0.04 & 606 & C\\
19.0 & 20.0 & 0.2 & 0.4 & -0.01 & 0.03 & 1 & 0.05 & 808 & D & -0.01 & 0.04 & 3 & 0.04 & 1461 & D\\
19.0 & 20.0 & 0.4 & 0.6 & 0.00 & 0.02 & 0 & 0.03 & 4122 & D & -0.02 & 0.03 & 0 & 0.05 & 9736 & D\\
19.0 & 20.0 & 0.6 & 0.8 & 0.00 & 0.03 & 2 & 0.04 & 901 & D & -0.01 & 0.03 & 1 & 0.05 & 977 & D\\
20.0 & 21.0 & 0.0 & 0.2 & -0.01 & 0.04 & 6 & 0.04 & 360 & E & -0.01 & 0.04 & 5 & 0.05 & 357 & E\\
20.0 & 21.0 & 0.2 & 0.4 & -0.02 & 0.03 & 2 & 0.05 & 695 & H & -0.02 & 0.04 & 7 & 0.05 & 867 & H\\
20.0 & 21.0 & 0.4 & 0.6 & -0.01 & 0.03 & 0 & 0.04 & 3885 & F & -0.02 & 0.04 & 1 & 0.07 & 5743 & F\\
20.0 & 21.0 & 0.6 & 0.8 & 0.00 & 0.03 & 1 & 0.04 & 2857 & F & -0.01 & 0.04 & 1 & 0.07 & 2426 & F\\
20.0 & 21.0 & 0.8 & 1.0 & 0.01 & 0.04 & 1 & 0.06 & 304 & F & 0.00 & 0.05 & 5 & 0.07 & 124 & F\\
21.0 & 22.0 & 0.0 & 0.2 & -0.02 & 0.04 & 6 & 0.04 & 396 & H & -0.02 & 0.04 & 6 & 0.05 & 401 & E\\
21.0 & 22.0 & 0.2 & 0.4 & -0.02 & 0.03 & 2 & 0.06 & 920 & E & -0.01 & 0.04 & 8 & 0.07 & 985 & H\\
21.0 & 22.0 & 0.4 & 0.6 & -0.01 & 0.03 & 1 & 0.06 & 5423 & F & -0.04 & 0.05 & 5 & 0.10 & 7517 & F\\
21.0 & 22.0 & 0.6 & 0.8 & -0.01 & 0.03 & 1 & 0.05 & 7147 & F & -0.02 & 0.05 & 2 & 0.12 & 6678 & F\\
21.0 & 22.0 & 0.8 & 1.0 & 0.00 & 0.04 & 1 & 0.07 & 2915 & F & -0.02 & 0.05 & 3 & 0.09 & 1317 & F\\
21.0 & 22.0 & 1.0 & 1.2 & 0.01 & 0.05 & 8 & 0.10 & 215 & F & 0.01 & 0.05 & 11 & 0.10 & 175 & F\\
22.0 & 23.0 & 0.0 & 0.2 & -0.02 & 0.04 & 7 & 0.05 & 306 & H & -0.02 & 0.04 & 4 & 0.05 & 294 & H\\
22.0 & 23.0 & 0.2 & 0.4 & -0.01 & 0.04 & 1 & 0.07 & 844 & H & 0.00 & 0.04 & 10 & 0.09 & 840 & H\\
22.0 & 23.0 & 0.4 & 0.6 & -0.01 & 0.04 & 1 & 0.07 & 4185 & F & -0.03 & 0.06 & 8 & 0.10 & 5517 & F\\
22.0 & 23.0 & 0.6 & 0.8 & 0.00 & 0.04 & 2 & 0.08 & 5584 & F & -0.03 & 0.06 & 4 & 0.16 & 7279 & F\\
22.0 & 23.0 & 0.8 & 1.0 & 0.01 & 0.05 & 1 & 0.09 & 5823 & F & -0.02 & 0.06 & 5 & 0.15 & 3021 & F\\
22.0 & 23.0 & 1.0 & 1.2 & 0.00 & 0.05 & 9 & 0.12 & 990 & F & 0.00 & 0.05 & 9 & 0.15 & 883 & F\\
22.0 & 23.0 & 1.2 & 1.4 & 0.06 & 0.05 & 32 & 0.14 & 204 & F & 0.04 & 0.04 & 31 & 0.15 & 179 & F\\
23.0 & 24.0 & 0.0 & 0.2 & -0.01 & 0.04 & 8 & 0.07 & 62 & H & -0.01 & 0.05 & 5 & 0.07 & 57 & H\\
23.0 & 24.0 & 0.2 & 0.4 & 0.00 & 0.04 & 4 & 0.08 & 428 & G & 0.02 & 0.05 & 13 & 0.64 & 291 & H\\
23.0 & 24.0 & 0.4 & 0.6 & 0.02 & 0.05 & 6 & 0.09 & 912 & H & 0.02 & 0.07 & 16 & 0.15 & 754 & H\\
23.0 & 24.0 & 0.6 & 0.8 & 0.02 & 0.05 & 10 & 0.15 & 765 & H & 0.00 & 0.07 & 16 & 0.20 & 1689 & H\\
23.0 & 24.0 & 0.8 & 1.0 & 0.02 & 0.06 & 6 & 0.12 & 2093 & H & 0.00 & 0.07 & 17 & 0.21 & 1604 & H\\
23.0 & 24.0 & 1.0 & 1.2 & -0.01 & 0.06 & 10 & 0.17 & 1057 & H & -0.01 & 0.06 & 10 & 0.22 & 942 & H\\
23.0 & 24.0 & 1.2 & 1.4 & 0.03 & 0.05 & 14 & 0.18 & 397 & H & 0.04 & 0.05 & 20 & 0.20 & 427 & H\\
	\enddata
	\tablecomments{We report quantities only for the bins where we have more than 50 galaxies.\\
$^\dag$: Spectroscopic survey from which the majority of the galaxies in the considered bin originates: A:SDSS/Galaxy Main Sample, B:E.Peng/AAT, C:E.Peng/Hectospec, D:SDSS/other programs, E:VVDS/F22, F:VIPERS, G:VVDS/F02, H:DEEP2/EGS.}
\end{deluxetable*}

\section{Angular correlation function}
\label{sec:woftheta}

A complementary way to test the photo-z accuracy using the whole NGVSLenS sample, is to calculate the galaxy angular correlation function, \citep[$w(\theta)$; e.g.,][]{newman08,hildebrandt09,mcquinn13}, in different redshift bins.
The advantage of this approach is that we probe the photo-z directly on the NGVSLenS data, and do not have to make assumption about the spectroscopic samples we are using.
This permits an estimation of the level of contamination between photometric redshift bins.
As a result of galaxy clustering, the angular correlation in a given redshift bin (auto-correlation) should be positive on small scales when compared to a random distribution of points; on large scales, the angular auto-correlation should tend to zero.
When looking at two redshift bins, the angular correlation (cross-correlation) should be zero if the redshift bins are well separated, since the galaxies are physically separated by large distances; if the considered redshift bins are close to each other and have sizes close to the typical photo-z uncertainty, this produces a non-zero cross correlation. 
We refer to \cite{erben09} for a detailed presentation of the angular correlation function formalism.

We apply the pairwise analysis using the publicly available \texttt{athena}\footnote{\href{http://cosmostat.org/athena.html}{http://cosmostat.org/athena.html}} tree code on our NGVSLenS data in redshift bins defined by the following limits:  $z_{cut} = 0.1;0.2; 0.3; 0.4; 0.5; 0.6; 0.7; 0.8; 0.9; 1.0; 1.2$.
Note that we neglect the effects of magnification \citep{scranton05,hildebrandt09a}.
We do not consider the range $0 \le z < 0.1$, as it contains too few objects (few thousands) to estimate robust statistics.
We only consider objects having $i < 23$ mag, not classified as star or GC (see Appendix~\ref{sec:StarGCrem}).
To exclude galaxies with unreliable $z_{\rm phot}$, we exclude galaxies with $z_{\rm phot,err.} > 0.25$ (according to Figure \ref{fig:zphoterr_ugriz}, the 3$\sigma$ upper limit of $z_{\rm phot,err.}$ at $i = 23$ mag is 0.25).
Also excluding all masked areas and field edges (to prevent duplicates when merging the fields), we end up with $4 \times 10^5$ objects when using photo-z's derived from $u^*griz$ bands, and $1.1 \times 10^6$ objects with just $u^*giz$ filters.
To compare with a random distribution, we generate random catalogs, having uniformly distributed positions with the same geometry as our NGVSLenS data (imaged areas and masks).
Because our chosen bin widths are at worse about twice as large than our photo-z scatter ($\sigma_{\rm outl.rej.} \lesssim 0.06$), we expect a low-level cross-correlation signal in neighboring  redshift bins, and a $w(\theta)$ compatible with zero for bins distant in redshift.

\begin{figure*}
	\includegraphics[scale=0.7]{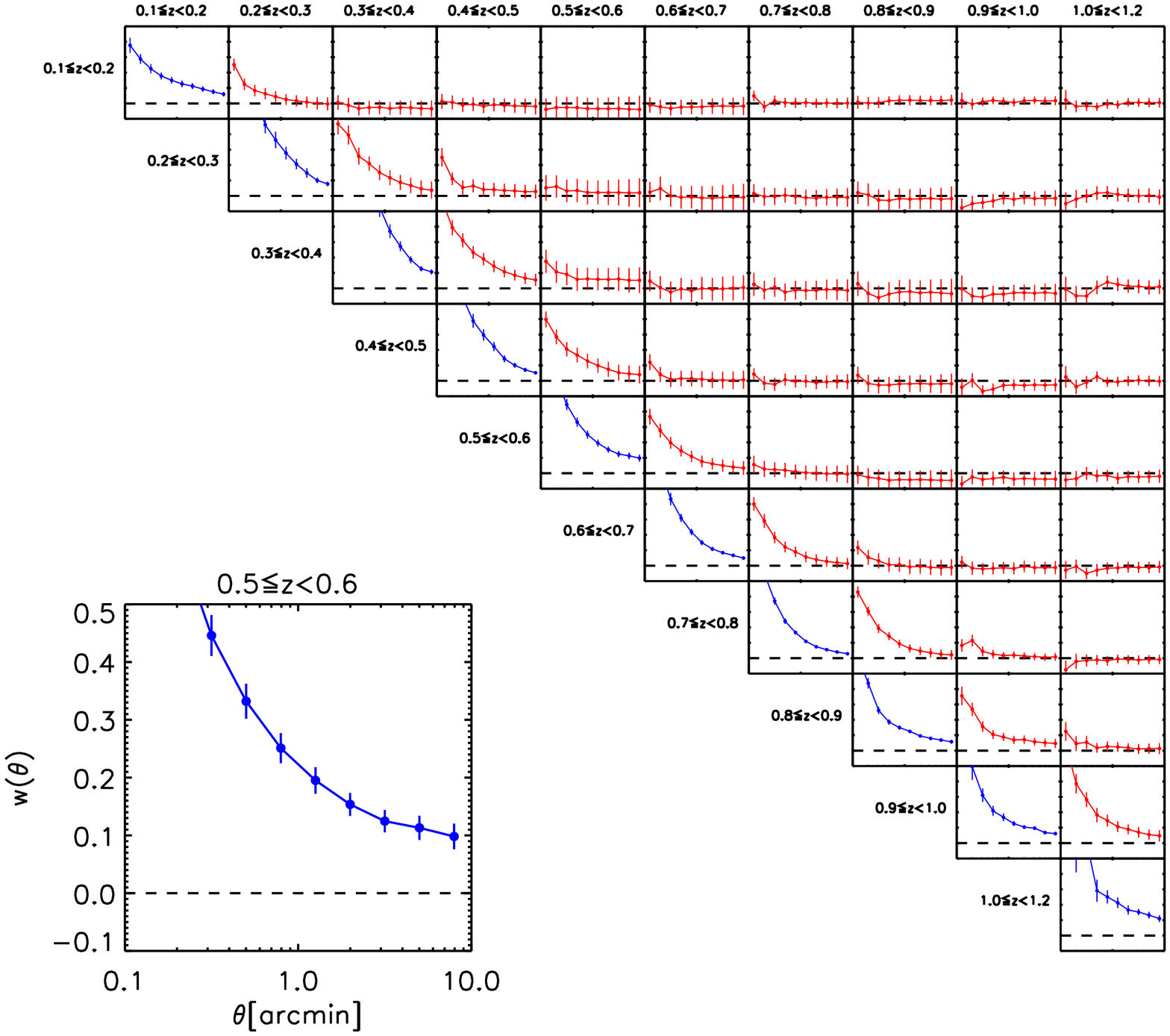}
\caption{Angular correlation for objects with $0.1 < z_{\rm phot} \le 1.2$, $i < 23$ mag, $z_{\rm phot,err.} \le 0.25$, not classified as star or GC, and covered by the $u^*rgiz$ bands, split by redshift bins.
Photo-z's have been estimated with \textit{Le Phare}.
The bottom left panel is a zoom-in to the corresponding panel of the matrix plot.
The absence of positive angular correlation in distant redshift bins is consistent with independent photometric redshift measurements.}
\label{fig:woftheta_ugriz}
\end{figure*}

Figure~\ref{fig:woftheta_ugriz} shows $w(\theta)$ for the photo-z's estimated with the $u^*griz$ bands and \textit{Le Phare}.
The figure -- hence the conclusions -- is similar if we use \textsc{bpz} or if we use the $u^*giz$ bands.
This figure is consistent with the expected behavior, which is a positive signal for auto-correlation (diagonal panels, in blue) and a signal consistent with zero for cross-correlation (red panels), except for adjacent redshift windows (second panel of each line, rightward of the blue panels) and for the second off-diagonal panels at small scales (third panel of each line).
A signal consistent with zero for cross-correlation in distant redshift bins confirms that our levels of contamination are minimal, as found in our previous analysis with spectroscopic samples.

\section{Conclusions}
\label{sec:conclusion}

We present an analysis of the determination of the photo-z catalog for the NGVS survey.
This survey images 104 deg$^2$ around the Virgo cluster with the $u^*giz$-bands, amongst which 34 pointings have $r$-band coverage.
To obtain good quality matched photometry, we used an upgraded version of the \textsc{theli} pipeline developed for the analysis of the CFHTLenS data, which have properties very similar to our data.
The \textsc{theli} pipeline products are the co-added astrometrically and photometrically calibrated images for all the NGVSLenS pointings in each filter, the photometric catalogs, and the photo-z catalogs.

We uniformly calibrated the photometry using the SDSS, which covers the full NGVSLenS. 
We built photometric catalogs from the multi--wavelength images convolved to the same seeing on each field.
This PSF homogenization allows an accurate measurement of colors, which is fundamental to obtain precise photo-z's.
We paid particular attention to the magnitude uncertainty estimation, accounting for the convolution process.
We estimate the photo-z's with two template fitting codes, \textit{Le Phare} and \textsc{bpz}.
We extended the prior of those codes to bright objects (seven magnitudes brighter), thus being able to accurately estimate photo-z's over a large range of magnitudes.

To assess the quality of our photo-z catalog, we used a large spectroscopic sample ($\sim$83,000 galaxies) in the  $0.01 \le z_{\rm spec} < 1.5$ and $15.5 \le i \le 24.5$ mag ranges.
We presented  a detailed analysis of our photo-z's as a function of the measured magnitude, redshift, and the number of bandpasses  used for their estimation ($u^*griz$ or $u^*giz$).

Our analysis concluded that both codes perform very similarly.
When using the $u^*griz$ bands, we obtain accurate photo-z's for $i \lesssim 23$ mag or $z_{\rm phot} \lesssim 1$: the bias is reasonable ($|bias| < 0.02$), the scatter $\sigma_{outl.rej}$ increases with the magnitude (from 0.02 to 0.05 for $i$ in [15.5,23]), and the outliers represent less than 5\% of the sample.
For photo-z's estimated without the $r$-band (i.e., with the $u^*giz$ bands), the accuracy decreases slightly.
The lack of $r$-band results in more pronounced uncertainties in the $0.3 \lesssim z_{\rm phot} \lesssim 0.8$ range, where it samples the 4,000 \AA~ break.
In this redshift range, we have $-0.05 < bias < -0.02$, $\sigma_{outl.rej} \sim 0.06$ and an outlier rate that peaks at 10-15\%.
The quality of photo-z's estimated with the $u^*giz$ bands also decreases at $i > 21$ mag.
However, we remark that  the brightest galaxies, e.g. the LRGs which constitute the main part of our NGVSLenS spectroscopic sample at $0.3 < z_{\rm spec} < 0.8$, have photo-z's of almost similar quality than when the $r$-band is used (but a slightly higher bias).
This is because of the different typical magnitudes of the LRGs with respect of the average galaxy in a given redshift bin: these galaxies have tighter priors because they are the brightest at a given redshift, and have a well defined 4,000 \AA~ break.
Our results are visualized and interpreted in a joint analysis in magnitude and redshift bins.

Finally, we presented an analysis of the angular correlation function $w(\theta)$, to internally assess the quality of our photo-z's using the whole NGVSLenS sample with $i \le 24$ mag and $0.1 \le z_{\rm phot} \le 2$.
We obtain results that are consistent with expectations, i.e.,  a positive signal for auto-correlation (decreasing with increasing angle) and a signal consistent with zero for cross-correlation when considering redshift bins with a redshift separation greater than 0.1.\\

The NGVSLenS catalogs will be public on June, 1st 2015 on the NGVS website\footnote{\url{https://www.astrosci.ca/NGVS/The_Next_Generation_Virgo_Cluster_Survey/Home.html}}.
Before that date, please contact us if you would like to use them \footnote{simona.mei@obspm.fr}.
 
\section*{Acknowledgments}
We thank the anonymous referee for his/her careful reading and suggestions, which improved the clarity of the paper.
The French authors acknowledge the support of the French Agence Nationale de la Recherche (ANR) under the reference ANR10-BLANC-0506-01-Projet VIRAGE.
SM acknowledges financial support from the Institut Universitaire de France (IUF).
H.H. is supported by the DFG Emmy Noether grant Hi 1495/2-1.
C.L. acknowledges support from the National Natural Science Foundation of China (Grant No. 11203017, 11125313 and 10973028).
R.P.M. acknowledges support from FONDECYT Postdoctoral Fellowship Project No.~3130750.
E.W.P. acknowledges support from the National Natural Science Foundation of China under Grant No. 11173003, and from the Strategic Priority Research Program, "The Emergence of Cosmological Structures", of the Chinese Academy of Sciences, Grant No. XDB09000105.
T.H.P. acknowledges support from FONDECYT Regular Grant (No. 1121005) and BASAL Center for Astrophysics and Associated Technologies (PFB-06).
H.Z. acknowledges support from the China-CONICYT Post-doctoral Fellowship, administered by the Chinese Academy of Sciences South America Center for Astronomy (CASSACA).
A.R. thanks Bego\~{n}a Ascaso for useful discussions.\\
 
This work is based on observations obtained with MegaPrime/MegaCam, a joint project of CFHT and CEA/DAPNIA, at the Canada--France--Hawaii Telescope (CFHT) which is operated by the National Research Council (NRC) of Canada, the Institut National des Sciences de lÕUnivers of the Centre National de la Recherche Scientifique (CNRS) of France and the University of Hawaii.
This research used the facilities of the Canadian Astronomy Data Centre operated by the National Research Council of Canada with the support of the Canadian Space Agency.
This publication has made use of data products from SDSS-III (full text acknowledgement is at \href{http://www.sdss3.org/collaboration/boiler-plate.php}{http://www.sdss3.org/collaboration/boiler-plate.php}).
Funding for the DEEP2 survey has been provided by NSF grants AST95-09298, AST-0071048, AST-0071198, AST-0507428, and AST-0507483 as well as NASA LTSA grant NNG04GC89G.
Some of the data presented herein were obtained at the W.M. Keck Observatory, which is operated as a scientific partnership among the California Institute of Technology, the University of California and the National Aeronautics and Space Administration.
The Observatory was made possible by the generous financial support of the W.M. Keck Foundation.
The authors wish to recognize and acknowledge the very significant cultural role and reverence that the summit of Mauna Kea has always had within the indigenous Hawaiian community. 
We are most fortunate to have the opportunity to conduct observations from this mountain. 
This research uses data from the VIMOS VLT Deep Survey, obtained from the VVDS database operated by Cesam, Laboratoire d'Astrophysique de Marseille, France. 
 This paper uses data from the VIMOS Public Extragalactic Redshift Survey (VIPERS).
 VIPERS has been performed using the ESO Very Large Telescope, under the "Large Programme" 182.A-0886.
 The participating institutions and funding agencies are listed at \href{http://vipers.inaf.it}{http://vipers.inaf.it}.


\appendix

\section{Prior extension to bright objects}
\label{sec:prior}

\textit{Le Phare} and \textsc{BPZ} were designed for high redshift studies.
Both codes use similar priors for $i>20$~mag galaxies (calibrated with $\sim$1,300 galaxies for \textsc{BPZ} and with $\sim$6,500 galaxies for \textit{Le Phare}), and for $i<20$~mag a prior that is not calibrated on observed data.
As a result of the large area covered by the NGVSLenS, $i<20$ mag galaxies represent a non-negligible fraction of our sample.
We thus build a new prior calibrated for bright \textit{and} faint objects.
Our estimated parameters are displayed in Table \ref{tab:prior}.

\begin{deluxetable}{l c c c c c c}
	\tablewidth{0pt}
	\tablecaption{Prior used in this article. \label{tab:prior}}
	\tablehead{
		\colhead{Spectral} & \colhead{$i_{ref}$} & \colhead{$\alpha_t$} & \colhead{$z_{0t}$} & \colhead{$k_{mt}$} & \colhead{$f_t$} & \colhead{$k_t$}\\
		\colhead{template type $T$} & \colhead{} & \colhead{} & \colhead{} & \colhead{} & \colhead{} & \colhead{}
		}
	\startdata
\multicolumn{7}{c}{$i_{AB} \le12.5$}\\
\hline
\multicolumn{7}{c}{$p(z) \propto 1$ if $z<0.1$, $p(z)=0$ else}\\
\hline
\multicolumn{7}{c}{$12.5<i_{AB}\le17$}\\
\hline
Ell			&	12.5		&	2.46		&	0		&	0.027	&	0.86	&	0.062\\
Spi			&	12.5		&	2.07		&	0		&	0.021	&	0.14	&	-0.108\\
Irr			&	12.5		&	1.89		&	0		&	0.015	&	...	&	...\\
\hline
\multicolumn{7}{c}{$17<i_{AB}\le20$}\\
\hline
Ell			&	17.0		&	2.46		&	0.121	&	0.103	&	0.65	&	0.257\\
Spi			&	17.0		&	1.94		&	0.095	&	0.098	&	0.23	&	-0.014\\
Irr			&	17.0		&	1.95		&	0.069	&	0.077	&	...	&	...\\
\hline
\multicolumn{7}{c}{$20<i_{AB}$}\\
\hline
Ell			&	20.0		&	2.46		&	0.431	&	0.091	&	0.30	&	0.40\\
Spi			&	20.0		&	1.81		&	0.390	&	0.100	&	0.35	&	0.30\\
Irr			&	20.0		&	2.00		&	0.300	&	0.150	&	...	&	...\\
\hline
	\enddata
\end{deluxetable}

According to the formalism introduced in \citet{benitez00} and using our template set (see Section \ref{sec:templates} and Table \ref{tab:photoz_setup}), for a galaxy with an $i$-band magnitude $i_{AB}$, the \textit{a priori} probability of having a redshift $z$ and a spectral template type $T$ is:
\begin{equation}
p(z,T | i_{AB}) = p(T | i_{AB}) \times p(z | T, i_{AB}).
\label{eq:prior0}
\end{equation}
$p(T | i_{AB})$ is the probability for a galaxy of magnitude $i_{AB}$ to have a spectral template type $T$ and is parametrized as:
\begin{equation}
p(T | i_{AB}) \propto f_t \times \textnormal{exp}\left( -k_t \times \left[i_{AB}-i_{ref}\right] \right).
\label{eq:prior1}
\end{equation}
$p(z | T, i_{AB})$ is the probability for a galaxy of magnitude $i_{AB}$ and spectral template type $T$ to have a redshift $z$ and is parametrized as:
\begin{equation}
p(z | T, i_{AB}) \propto z^{\alpha_t} \times \textnormal{exp}\left( -\left[ \frac{z}{z_{0t}+k_{mt}\times(i_{AB}-i_{ref})} \right]^{\alpha_t} \right).\\
\label{eq:prior2}
\end{equation}
$i_{ref}$ is a reference magnitude.

For galaxies with $i>20$ mag, we use the \textit{Le Phare} prior, calibrated on the robust VVDS spectroscopic sample.
For galaxies with $12.5<i\le17$ mag, we calibrate our prior with the SDSS spectroscopic Galaxy Main Sample \citep{york00,strauss02}, using the DR10 release \citep{ahn14}.
We select objects with \texttt{class$=$GALAXY} and use \texttt{cModelMag\_i} as $i$-band total magnitude and \texttt{ModelMag} quantities to compute colors (we correct for extinction).
This sample is complete down to $r \le 17.77$ mag, thus complete down to $i \sim17$ mag (in fact, most of the galaxies with $r \le 17.77$ mag have $z\lesssim0.4$, and for $z\lesssim0.4$ the color $(r-i)\lesssim0.8$; see Figure~\ref{fig:template_color}).
For consistency with standard prior already implemented in \textsc{BPZ} and \textit{Le Phare}, we follow the formalism of \citet{benitez00} and define three broad spectral classes: Ellipticals (Ell template), Spirals (Sbc, Scd templates), and Irregulars (Im, SB2, SB3 templates).
We associate each galaxy from the SDSS spectroscopic Galaxy Main Sample to one of those broad spectral classes, by using the best-fit template (from the photometric redshift code) when fixing the redshift at $z_{\rm spec}$.
We thus have $\sim$320,000 galaxies with $12.5<i<17$~mag and classified into three broad spectral classes.
We then fit with a least-square fitting method \citep[IDL \textsc{MPFIT} package,][]{markwardt09}  the fraction of each spectral type as a function of the magnitude with equation \ref{eq:prior1} and we obtain the $k_t$ parameter, and the redshift distribution of each spectral type T in each magnitude bin with equation \ref{eq:prior2} and obtain the $\alpha_t$, $z_{0t}$, $k_{mt}$ parameters.
For $17<i\le20$ mag galaxies, we extrapolate the parameter values to match those fitted at $i_{AB}=17$ mag and $i_{AB}=20$ mag.
For $\alpha_t$, we take the mean value of those estimated at $i_{AB}=17$ mag and $i_{AB}=20$ mag, and we set the other parameters so that the quantities $z_{0t}+k_{mt}\times(i_{AB}-i_{ref})$ and $f_t \times \textnormal{exp}(i_{AB}-i_{ref})$ are continuous.
For $i\le12.5$ mag galaxies, we set a square prior, $p(z)$ being non-null for $z<0.1$ and null for $z\ge0.1$.\\

\begin{figure}[!h]
	\centering
	\includegraphics[width=0.5\columnwidth]{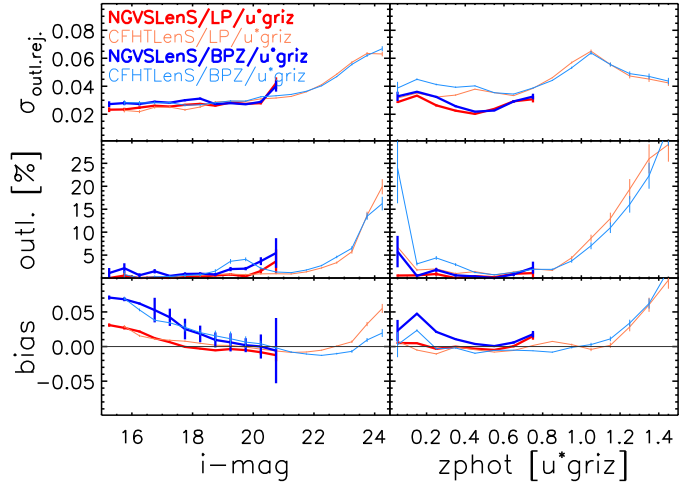}\\
	\caption{
Statistics for photo-z's (estimated with $u^*griz$ bands) as a function of magnitude (\textit{left panel}) and redshift (\textit{right panel}), when using \textit{Le Phare} and \textsc{BPZ} original priors.
Symbols are similar to those in Figure~\ref{fig:stats_ugriz}.
At $i < 20$ mag, the bias is larger than when using our re-computed prior in Figure~\ref{fig:stats_ugriz}.}
\label{fig:PRIOR_ORIG_stats_ugriz}
\end{figure}

To illustrate how our new prior improves the photo-z accuracy, we display in the right panel of Figure~\ref{fig:PRIOR_ORIG_stats_ugriz} the statistics for the photo-z estimated with $u^*griz$-bands and using \textit{Le Phare} and \textsc{BPZ} original priors.
When comparing with the Figure~\ref{fig:stats_ugriz}, we see that our new prior, for $i < 20$ mag galaxies, improves significantly the bias and, to a lesser extent, the scatter and the percentage of outliers; as a consequence, it also improves the photo-z accuracy at $z_{\rm{phot}} < 0.4$.

\section{Similarity of the NGVSL\lowercase{en}S and CFHTL\lowercase{en}S datasets}
\label{sec:lensprop}

In Section~\ref{sec:zphotqual}, we assume that the CFHTLenS dataset -- re-reduced with our changes in the \textsc{theli} pipeline -- has similar properties as our NGVSLenS datasets.
This is justified by the fact that both datasets have been obtained with the same telescope, instrument, and filterset, have by construction similar depth, and have been reduced with the same pipeline.

\begin{figure}[!h]
	\centering
	\includegraphics[scale=0.5]{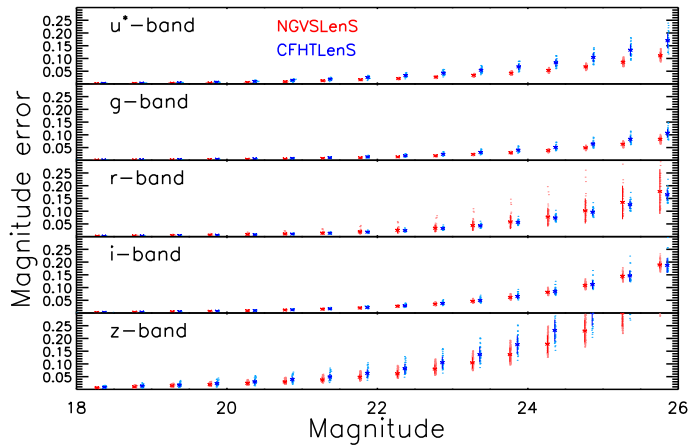}
	\caption{Magnitude uncertainty as a function of magnitude for the NGVSLenS (red) and the CFHTLenS (blue) data, and for the $u^*griz$ bands.
Each point represent the median values for a given field (i.e.,  1 deg$^2$).
The crosses and error bars represent the global values for each survey (median and standard deviation of those individual -- i.e.,  for each field -- data).
CFHTLenS data are offset by 0.1 mag along the x-axis to increase readability.
Both datasets have very similar depths for all bands.}
\label{fig:lensprop}
\end{figure}

In this Section, we illustrate, \textit{a posteriori} the similarity of the datasets resulting from these two surveys.
We display in Figure~\ref{fig:lensprop} our estimated uncertainties in magnitude as a function of magnitude for both datasets.
Each point represents the median value of the magnitude error for a given magnitude bin in each field.
The crosses and error bars represent the global values for each survey (median and standard deviation of those individual -- i.e.,  for each field -- data).

We observe that two datasets have indeed very similar depths for all bands.
The NGVSLenS has slightly better photometry in the $u^*$- and $z$-bands.
We note the large standard deviations for the $r$-band for the NGVSLenS, that is due to the continuation of the NGVSLenS program for some fields with smaller exposure times \citep[see Section 4.4 of][]{ferrarese12}.

\section{Stars and Globular clusters removal}
\label{sec:StarGCrem}

In the Figures using photometric samples, we exclude objects that are most probably stars or GCs.
We describe in this Appendix the criteria we use to exclude those objects.
For more detailed NGVS studies of stars and GCs, please refer to Lokhorst et al. (\textit{in preparation}) and \citet{durrell14}, respectively.

\subsection{Stars removal}

To separate stars from galaxies we follow a simple approach.
We classify an object as a star if it satisfies:
$\texttt{CLASS\_STAR} > 0.95$ and $i < 21.5$ mag, where \texttt{CLASS\_STAR} is a \textsc{SExtractor} output quantifying the star/galaxy classification.
To assess the efficiency of our classification, we have in hand $\sim$6,900 CFHTLenS spectroscopically identified stars ($i = 20.7 \pm 1.4$ mag;  using the DEEP2/EGS, VVDS, and VIPERS) and $\sim$3,000 NGVSLenS spectroscopic stars ($i = 18.9 \pm 1.5$ mag; using the SDSS), and our galaxy spectroscopic samples (NGVSLenS and CFHTLenS; $\sim$83,000 galaxies with $0.01 \le z_{\rm spec} < 1.5$).

\begin{figure}[!h]
	\centering
	\includegraphics[scale=0.5]{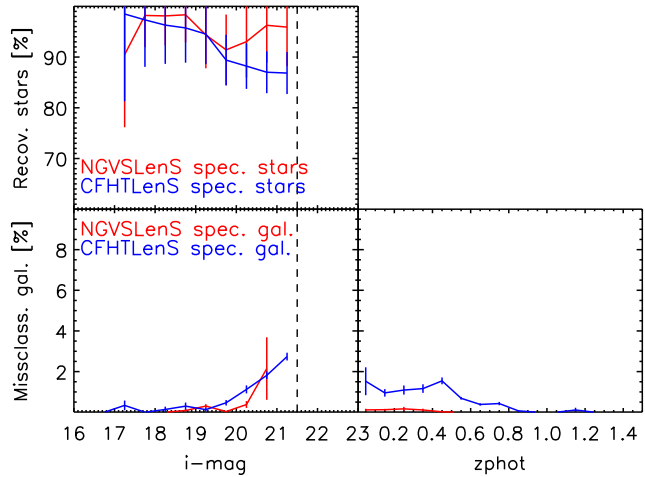}
	\caption{Star/galaxy separation efficiency as a function of the $i$-band magnitude (left panel) or $z_{\rm phot}$ (right panel).
The upper panel represent the percentage of stars correctly classified as stars; the lower panels represent the percentage of galaxies wrongly classified as stars.
Please see text for description of the samples.
The vertical dashed lines show the faint limit ($i = 21.5$ mag) we use to assign the star classification.
We report quantities only for the bins where we have more than 50 objects.
Error bars are calculated assuming a Poissonian distribution.}
\label{fig:SGclass}
\end{figure}

We present in Figure~\ref{fig:SGclass} the percentage of stars correctly classified as stars (upper panel) and the percentage of galaxies misclassified as stars (bottom panel) when using our criteria.
We correctly classify $>85\%$ of the stars with $i<21.5$ mag and misclassify as stars $<5\%$ of our galaxies.
Note that, at $i = 22$ mag, we expect galaxies to be about 10 times more numerous than stars \citep[see Figure~3 of ][for the COSMOS field, which lies at a Galactic latitude lower than the NGVSLenS]{lilly07}, and even more at fainter magnitudes.
Therefore, for $i > 21.5$ mag objects, the number of  stars should be marginal when compared to the number of galaxies and should not affect the statistics.
The performance of our star/galaxy classification is thus satisfactory for our needs.

\subsection{Globular clusters removal}

A special feature -- and the main goal -- of the NGVS is its full coverage of the Virgo cluster: a direct consequence is a massive detection of Globular Clusters (GCs).
While this provides a unique sample for Virgo science, these GCs may contaminate \textit{background-science} samples: in fact, as they are relatively small, red, and faint objects, they can easily be mistaken for higher redshift objects.

We provide here the criteria we used to exclude GCs from our (photometric) samples in this study.
The requirements, depending on the galaxy populations under study, is a trade-off between maximizing the number of GCs removed and minimizing the number of background sources removed.
For this paper, as we are not studying a special galaxy population, we aim at removing a significant portion of the GCs while accepting the loss of a marginal number of background sources.

To test our method, we use our spectroscopic samples (NGVSLenS and CFHTLenS) with $0.01 \le z_{\rm spec} < 1.5$ ($\sim$83,000 galaxies) for background sources, and a sample of $\sim$750 confirmed GCs around M87 \citep{hanes01,strader11}.
As detailed in Section~\ref{sec:samples}, our spectroscopic sample, being composite, fairly covers the different galaxy populations up to $z_{\rm spec} < 1.5$.
After several tests, the following criteria gave the best results (according to our requirements): for removing GCs:
(1) $\texttt{CLASS\_STAR} > 0.05$,
(2) $18.0 < \texttt{MU\_MAX}\footnote{\texttt{MU\_MAX} is a \textsc{SExtractor} output measuring the peak surface brightness.} < 21.5$,
(3) $1.5 < u-z < 3.0$,
(4) $z_{\rm phot} < 0.2$.

Applying these criteria results in the removal of 94\% of the GCs and  $0.2\%$ of our spectroscopic sample (the vast majority of misclassified galaxies have $z_{\rm spec} < 0.25$).
We note that these results do not depend on whether the photo-z's are estimated with \textit{Le Phare} or \textsc{bpz}, or with the use of $u^*griz$ or $u^*giz$ bands.
Our criteria to remove GCs also remove some stars (e.g., it removes 6\% of our CFHTLenS spectroscopic stars), and \textit{vice-versa}.
Although we are not able to properly separate stars and GCs, we are able to efficiently remove the vast majority of stars and GCs with a marginal loss of galaxies.\\

\end{document}